\documentclass[aps,prd,showpacs,showkeys,preprintnumbers,floatfix,nofootinbib]{revtex4-2}
\usepackage[utf8]{inputenc}
\usepackage{graphicx}
\usepackage{amsmath}
\usepackage{tikz}
\usepackage{amsfonts} 
\usepackage{physics}
\usepackage{qcircuit}
\usepackage{comment}
\usepackage{hyperref}
\hypersetup{
     colorlinks   = true,
     citecolor    = blue
}

\usepackage{bbm}
\DeclareMathAlphabet{\mymathbb}{U}{bbold}{m}{n}

\newcommand{\negspace}{\!}
\newcommand{\lsub}[2]{{\protect\vphantom{#1}}_{#2} \negspace {#1}}
\newcommand{\rsub}[2]{{#1} \negspace {\protect\vphantom{#1}}_{#2}}

\newcommand{\ketsub}[2]{\rsub {\ket{#1}} {#2}}

\newcommand{\brasub}[2]{\lsub {\bra{#1}} {#2}}
\newcommand{\braketsub}[4]{\rsub {\lsub {\braket{#1}{#2}} {#3}}{#4}}

\newcommand{\nn}{\nonumber}
\usepackage{color}
\definecolor{jlab_red}{RGB}{192,39,45}
\definecolor{jlab_orange}{RGB}{249,102,0}
\definecolor{jlab_blue}{RGB}{47,122,121}
\definecolor{jlab_green}{RGB}{65,125,10}

\def\be{\begin{equation}}
\def\ee{\end{equation}}
\def\bea{\begin{eqnarray}}
\def\eea{\end{eqnarray}}

\usepackage{ulem} 
\usepackage{bm}

\newcommand\eq[1]{Eq.~\eqref{eq:#1}}

\newcommand\eqs[2]{Eqs.~\eqref{eq:#1} \& \eqref{eq:#2}}
\newcommand\eqlist[2]{Eqs.~\eqref{eq:#1}-\eqref{eq:#2}}

\newcommand\fig[1]{Fig.\ \ref{fig:#1}}

\newcommand\secn[1]{Section~\ref{secn:#1}}

\newcommand{\qwred}[1][-1]{\ar @[red]@{-} [0,#1]}

\newcommand*{\UCB}{Department of Physics, University of California, Berkeley, CA 94720, USA}    
\newcommand*{\LBNL}{Nuclear Science Division, Lawrence Berkeley National Laboratory, Berkeley, CA 94720, USA}  \newcommand*{\Jlab}{Thomas Jefferson National Accelerator Facility, 12000 Jefferson Avenue, Newport News, Virginia 23606, USA}    
\newcommand*{\UVA}{Department of Physics, University of Virginia, 382 McCormick Rd, Charlottesville, Virginia 22904-4714, USA}    
\newcommand*{\UTK}{Department of Physics and Astronomy, University of Tennessee, Knoxville, Tennessee 37996-1200, USA}

\begin{document}
\title{
Toward coherent quantum computation of scattering amplitudes \\
with a measurement-based photonic quantum processor
}

\author{Ra\'ul A.~Brice\~no}
\email[e-mail: ]{rbriceno@berkeley.edu}
\affiliation{\UCB}\affiliation{\LBNL} 

\author{Robert G.~Edwards}
\email[e-mail: ]{edwards@jlab.org}
\affiliation{\Jlab}


\author{Miller Eaton}
\author{Carlos Gonz\'alez-Arciniegas}
\author{Olivier Pfister}
\email[e-mail: ]{olivier.pfister@gmail.com}
\affiliation{\UVA}


\author{George Siopsis}
\email[e-mail: ]{siopsis@tennessee.edu}
\affiliation{\UTK}
\date{\today}
\begin{abstract}

In recent years, applications of quantum simulation have been developed to study properties of strongly interacting theories. This has been driven by two factors: on the one hand, needs from theorists to have access to physical observables that are prohibitively difficult to study using classical computing; on the other hand, quantum hardware becoming increasingly reliable and scalable to larger systems. In this work, we discuss the feasibility of using quantum optical simulation for studying scattering observables that are presently inaccessible via lattice QCD and are at the core of the experimental program at Jefferson Lab, the future Electron-Ion Collider, and other accelerator facilities. We show that recent progress in measurement-based photonic quantum computing can be leveraged to provide deterministic generation of required exotic gates and implementation in a single photonic quantum processor.

\end{abstract}

\maketitle 

\section{Introduction}

It is now well-accepted that the vast majority of subatomic phenomena are described by the Standard Model of particles. This encodes three of the known fundamental forces, the opaquest of which is the strong nuclear force. This force is responsible for the formation and structure of all atomic nuclei and countless of other experimentally observed particles.\footnote{Collectively, asymptotic states of QCD are known as hadrons.} Although we know this force is described by quantum chromodynamics (QCD), the theory of quarks and gluons, it is not yet evident how the dynamics of these fundamental particles result in emergent phenomena taking place throughout the universe and experiments across the world. As a result, one of the primary goals of modern-day nuclear physics has been to construct accurate methods for accessing the consequences of the theory. 

The primary challenge in studying QCD is its non-perturbative nature. This has motivated the development of a large variety of frameworks to constrain physical quantities. Arguably, the most successful method that has been demonstrated to be able to access QCD observables without making approximations on the underlying dynamics goes by the name of lattice QCD.  Lattice QCD makes use of Monte Carlo techniques to statistically sample the Euclidean path integral of QCD. This provides statistical determination of correlation functions of the theory defined in Euclidean spacetime, which can subsequently be related to a variety of key observables of the theory. Most notably, one can determine quantities that are independent of time, like the spectrum and local matrix elements.\footnote{We point the reader to Refs.~\cite{Aoki:2021kgd} and \cite{Briceno:2017max} for recent reviews on the progress of lattice QCD community to measure single- and multi-hadron observables respectively.}

As one would expect, this procedure puts tight constraints on the physics that may be directly accessed. In particular, quantities that are sensitive to the time signature or static properties of systems with finite chemical potential are currently just outside of the reach of lattice QCD calculations. In general, these quantities introduce sign problems that prohibit their direct calculation.\footnote{Some methods have been constructed to indirectly access time-dependent quantities [e.g.,  Refs.~\cite{Hansen:2017mnd,Bulava:2019kbi}].}  
The desire to have real-time, non-perturbative quantities, among other observables that can not be obtained directly via lattice QCD, has turned the field towards considering the use of quantum computing (QC).

 A driving motivation for having a non-perturbative real-time formulation of quantum field theories (QFT) is the determination of scattering amplitude of few-body systems. These can obtained from infinite-volume real-time correlation functions following the Lehmann, Symanzik and Zimmermann  reduction procedure. Most physical properties of few-body systems are encoded in scattering amplitudes and they are the primary focus of particle and nuclear experimental searches. As a result, having a non-perturbative mechanism to access these from the Standard Model is of utmost importance. In this work, we focus our attention on the feasibility of using quantum optical simulation for determining scattering observables for QFTs. We begin by briefly reviewing the history of this problem.

As first remarked by Feynman~\cite{Feynman1982}, using QC for the quantum simulation 
of an $N$-qubit physical system provides an exponential speedup because QC requires but a polynomial overhead $\sim\mathcal O(N^p)$ (accounting for quantum error correction) whereas classically solving  quantum evolution, e.g., Schr\"odinger's equation, requires diagonalizing a $2^{2N}$-dimensional Hamiltonian matrix. Lloyd confirmed this quantum advantage~\cite{Lloyd1996} by remarking that a vast majority of physically relevant Hamiltonians (hard-sphere and van der Waals gases, Ising and Heisenberg spin systems, strong and weak interactions, and some gauge theories) have a local structure, yielding a polynomial number $\sim\mathcal O(N^k)$ of Hamiltonian parameters. Building on these
seminal works, 
a great deal of progress has been made in setting a pathway towards the quantum simulation of quantum field theories~\cite{Jordan2012, Jordan2014, Marshall2015a, Briceno:2020rar, deJong:2021wsd, DeJong:2020riy, Bauer:2021gek, Nguyen:2021hyk, Andrade:2021pil,Davoudi:2021ney}. 
See also Ref.~\cite{PRXQuantum.4.027001} for a recent whitepaper on the topic. 

Whereas quantum evolution in QS can be implemented by ``brute force'' QC  using the gate/circuit model, a potentially more promising approach is to find direct mappings between the physical system to be simulated and the experimental simulator. An exemplar is the quantum simulation of the quantum field theory of a self-interacting scalar, initially proposed over qubits~\cite{Jordan2012,Jordan2014} and subsequently adapted to continuous-variable quantum optical fields, a.k.a.\ qumodes, in lieu of qubits, in Ref.~\cite{Marshall2015a}. In both cases, an effective field theory is employed with spatial discretization of the field. In the latter case, an additional benefit is the simulation of massive bosonic quantum fields with  quantum optical fields.

Subsequently, Ref.~\cite{Briceno:2020rar} raised the issue that the need to restrict the space to be finite would na\"ively prohibit the access to scattering observables, given the absence of asymptotic states in a finite volume. This same article provided a practical solution to this issue, by constructing carefully-defined wavepackets, thereby allowing physical scattering amplitudes to be, in principle, determined with finite resources.   

The ability to ultimately perform any quantum simulation hinges upon reliable quantum hardware. In recent years there has been some progress on this front.  Two approaches might be defined: either find a breakthrough application for a noisy intermediate-scale quantum (NISQ) device~\cite{Preskill2018} that would yield genuine quantum advantage, or find a path beyond the NISQ regime to achieve scalable, fault-tolerant quantum computing. While the former approach is applicable to simpler quantum technology such as quantum sensing (e.g., LIGO~\cite{Aasi2013}), it has not yet been shown to be relevant to quantum computing (QC).

More specialized, non-universal quantum machines, such as quantum circuit samplers~\cite{Arute2019}, boson samplers~\cite{Aaronson2010}, and Gaussian boson samplers~\cite{Hamilton2017}, have claimed a quantum advantage over classical computers for sampling from a hard-to-calculate probability distribution~\cite{Arute2019,Zhong2020,Madsen2022}. 
Quantum samplers operate efficiently by simply measuring the outputs of their quantum circuit, whose unitary evolution is hard to calculate classically, the measurement results constituting samples from the associated classically intractable quantum probability distribution.

A useful application for sampling would be to reconstruct intractable probability distributions from statistically significant samples. This was proposed, for example, to compute Franck-Condon factors in molecular spectra~\cite{Huh2015} or to detect graph isomorphism~\cite{Bradler2021}. 

However, these proposed extensions of boson sampling beyond its initial intent do face a major setback that negates their promise: the number of different measurement outcomes needing to be independently sampled typically suffers an exponential, or even super-exponential, increase with the number of qubits or qumodes in the quantum circuit. This is why the quantum advantage demonstrated by quantum samplers so far~\cite{Arute2019,Zhong2020,Madsen2022} remains a sampling advantage, not a computational one.

It is also expected to be the case that sampling the complete distribution of states of a given QFT is an exponentially hard problem. That said, our goal is to provide a method for computing scattering processes using quantum processors. As we will argue in Sec.~\ref{secn:obs}, all scattering observables can be reconstructed from time-dependent correlation functions of just single-particle states in the theory. This implies that one only needs to access such states, and this may be the key to alleviating the complexity of the distribution problem.   

In this work, we derive the various quantum optical building blocks necessary to quantum simulate time-dependent amplitudes for scalar field theories. In particular, we focus our attention on the determination of matrix elements involving time-separated currents.  In Sec.~\ref{secn:obs}, we begin by providing a detailed description of the classes of matrix elements we are after. Furthermore,  provide a conceptually simple for accessing spectra and matrix elements of states of an interacting theory using real-time correlation functions. In Sec.~\ref{secn:2}, we discuss a procedure for using quantum optics for performing quantum simulations of complex field theory. In \secn{timephase}, we apply this procedure to the  \textit{coherent} quantum simulation of correlators. Finally, in \secn{mbqc}, we show how techniques from measurement-based quantum computing can be leveraged to simulate current correlators.
 
\section{Physical observables from Minkowski correlators}\label{secn:obs}

We begin by reviewing a class of scattering amplitudes that are at the core of the experimental program at Jefferson Lab and the future Electron-Ion Collider, but that  are presently inaccessible via lattice QCD. These are time-sensitive matrix elements of the form $\langle P_f | \mathcal{J}^\mu (t) \mathcal{J}^\nu (0)|P_i \rangle $, where $t$ is real time separating the electromagnetic current $\mathcal{J}^\mu$, $\mu$ is the Lorentz index, and $|P_i\rangle$ and $\langle P_f|$ denote generic initial and final states with four-momenta $P_i$ and $P_f$ respectively. By determining these matrix elements as a function of time, one can obtain a rich set of the structure of information of the initial/final states. As a concrete example, if one considers the case where $P_f^2=P_i^2=m^2$, where $m$ is the mass of a stable hadron, the Fourier transform of this matrix element results in the virtual Compton scattering amplitude, 
\begin{align}
\mathcal{C}
=i\int dt \, d\vec{x} \, e^{i q \cdot x} \langle P_f | \mathcal T\big[\mathcal{J}^\mu (t,\vec{x}) \mathcal{J}^\nu (0)\big]|P_i \rangle_{\rm conn.},
\label{eq:Compton}
\end{align}
where we have left the kinematic dependence of the amplitude implicit, the ``conn." subscript means that only fully connected diagrams contribute to the amplitude, $q$ is the four-vector denoting the momenta carried by one of the external currents, we have assumed the states have appropriate relativistic normalization, and $\mathcal T$ denotes the time-ordering operator.\footnote{A detailed recent discussion of this amplitude can be found in Refs.~\cite{Sherman:2022tco, Briceno:2019opb}.} This virtual Compton scattering amplitude can be used, for example, to constrain the basic functions describing the distributions of the quarks and gluons inside the hadron~\cite{Ji:1996nm}.

This is just one example of a broad class of amplitudes that are currently inaccessible via Euclidean correlation functions that may, in principle, be determined using Minkowski correlators. More generally, by determining matrix elements of time-dependent products of external currents, $\langle P_f | \prod_{n} \mathcal{J} (x_n) |P_i \rangle$, and using the Lehmann-Symanzik-Zimmermann reduction formula one can quite literally obtain any scattering amplitude desired~\cite{Carrillo:2023}. This would be a major advantage over lattice QCD. Here we will discuss at most the first non-trivial example, namely amplitudes of the same class as the Compton scattering amplitude.

Having given a concrete example of an observable of interest, it is worthwhile to make some generalities regarding the access of scattering amplitudes via numerical methods. It is generally true that physical quantities of a QFT can be obtained directly or indirectly from real-time correlation functions. In practice, in order to evaluate correlation functions, one must resort to introducing regulators. Lattice field theories are defined by the introduction of specific ultraviolet (UV) and infrared (IR) regulators. The UV regulator is the smallest allowed spacetime separation, known as the lattice spacing, and is normally labeled as $a$. In contrast, the IR regulator is the size of the spacetime volume ($\mathcal{V}$). For example, if we consider theories in 1+1D, the spacetime volume would be $\mathcal{V}=T\times L$ where $T$ is the temporal extent and $L$ is the spatial extent. Na\"ively, one can recover a desired physical observable $\mathcal{M}$ by taking the limit in which these regularities are removed, e.g., $\mathcal{M} = \lim_{a\to 0, L\to\infty}\mathcal{M}(a,L)$. In general, these limits are either not well defined  or impractical to achieve. This subtle point is of significant importance when interested in the determination of scattering amplitudes. As previously mentioned, Ref.~\cite{Briceno:2020rar} explained why scattering amplitudes are not directly accessible from correlation functions defined in a finite space due to the absence of asymptotic states. That being said, this limitation is lifted in the aforementioned reference. 

\subsection{Definitions and notations}

As discussed in the introduction, a primary shortcoming of the modern-day lattice QCD program is its inability to access correlation functions defined in real time. Here we discuss how time-dependent matrix elements, of the kind shown on the right-hand side of \eq{Compton}, could in principle be obtained using real-time quantum computations of the theory. To access these, it will be necessary to first consider the simpler correlation functions. 

Before proceeding with the definition and understanding of the desired correlation functions, it is useful to introduce some concepts. First, for simplicity, we will assume that the fields of the underlying quantum field of interest are real scalars. This simplifies the subsequent expressions while leaving the key message unchanged.

Second, we will assume the presence of an IR regulator that renders the spectrum discrete. A natural example of this would be making the volume finite, which will be necessary in future calculations. This would also make the momenta discrete.  

Such a regulator would discretize the eigenstates of both the free and full Hamiltonian labeled as $H_0$ and $H$, respectively, with eigenstates and eigenvalues defined by 
\begin{align}
    H_0  \ketsub{\ell}{o} 
&= E^{(0)}_{\ell} \ketsub{\ell}{o}, \\
H \ket{\ell}
&= E_{\ell} \ket{\ell}, \label{eq:H} 
\end{align}
where $\ell$ is a vector comprised of the quantum numbers that define the eigenstate.
We reserve the special symbol of $\ketsub{\Omega}{o}$ for the vacuum of the free Hamiltonian ($\ketsub{\Omega}o = \ketsub 0 o$), which we conveniently fix to have zero energy ($E_0^{(0)} = 0$). Note that we will use the state normalization
\begin{align}
    \braket{\ell_f}{\ell_i}=\delta_{\ell_i,\ell_f},
\end{align} which differs from the standard infinite-volume normalization for relativistic states.

To simulate dynamics, we will use optical field modes and identify the eigenstates of the free Hamiltonian $\ketsub n o$ with the Fock states of the optical simulating field,
\begin{equation}\label{eq:4}
    N\ketsub{n}\gamma=n\ketsub{n}\gamma \ ,
\end{equation}
where $N=a^\dag a$ is the photon number operator of the corresponding mode, and the subscript $\gamma$ emphasizes that the state is defined in terms of the optical field.
We will also need the eigenstates of the amplitude and phase quadratures of the quantum optical field (see \eq E in the Appendix), respectively,
\begin{align}
    Q\ketsub s {\gamma_q} &= s\ketsub s{\gamma_q}\nonumber\\
    P\ketsub s {\gamma_p} &= s\ketsub s{\gamma_p}\ .
\end{align}
The quadratures $Q$ and $P$ act similarly to the position and momentum, respectively, of a harmonic oscillator.

We will concentrate on matrix elements of currents, which are operators that can be written in terms of the fundamental fields of the theory, and their matrix elements can encode the physical properties of the states. A notable example is the electromagnetic vector current, $\mathcal{J}^\mu$ whose matrix elements can, among other things, provide constraints on the charge distribution inside a given state,  as well as other observables of interest.

\subsection{Accessing the spectrum }
Accessing any one state of the interacting theory is generally challenging. A more modest goal is to probe properties of these states. Here we introduce a methodology for accessing the spectrum of the full Hamiltonian, without having to resolve the eigenstates.\footnote{Note that the former is compatible with quantum simulation whereas the latter is not, as it carries the well-known exponential overhead of quantum state estimation.} Instead, we use eigenstates of the free Hamiltonian, which are more readily accessible as probes. This is a soft assumption since one could use any basis of states as probes into the eigenstates of the full Hamiltonian, as long as these overlap with the desired sector of the theory. 

Let $|\ell_i\rangle_o$ and $|\ell_f\rangle_o$ be arbitrary eigenstates of $H_0$, where $\ell_i$ and $\ell_f$ are vectors comprised of the quantum numbers defining the corresponding states. Then we define a one-point correlation function, 
\begin{align}
\label{eq:Ct}
 C_{1 \rm pt.}(t) 
&= \brasub{\ell_f}o  e^{-it H} \ketsub{\ell_i}o.
\end{align}
 as a matrix element of the unitary evolution operation on the basis of free states of the theory, making it suitable for quantum simulation with the free states identified with suitable Fock states of the optical field. 

By inserting a complete set of states, we can rewrite the correlation function as
\begin{equation}
 C_{1 \rm pt.}(t) 
= \sum_{n}
 e^{-it E_n} 
\braketsub{\ell_f}no{}
\,\braketsub n {\ell_i}{}o
\ ,
\label{eq:C1pt_spec}
\end{equation}
making explicit the connection between the time dependence of the correlators and the spectrum of the full Hamiltonian $H$.

Modern-day lattice QCD calculations exploit a similar relation between Euclidean correlators and the spectrum, where the former can be written as $\sum_n c_{n} e^{- \tau E_n}$ in imaginary time, $\tau$. The coefficients  $c_{n}$ encode the overlap between desired states and the operators used to access them. Since the spectrum is time-independent, the spectra obtained using Minkowski and Euclidean correlation functions are identical. 
                                
Since Euclidean correlators are sums over exponentially decaying functions, they are dominated by low-lying states at ``\emph{large}" imaginary time, allowing systematic determination of these states. Determining the energies of the theory from Minkowski correlators is slightly more challenging since all states in principle contribute at any given time. Various techniques have been proposed in the literature~\cite{Qian:2021wya, Choi:2020pdg, Kaplan:2017ccd}. Here we present a conceptually simple but perhaps computationally impractical one. As will be shown, in addition to being systematically improvable, the advantage of this method is that it is easily generalizable to more complicated quantities like the desired long-range matrix elements.     

Assuming we have constrained the correlation functions for times in the range of $t \in [0,T]$, \eq{C1pt_spec} can be Fourier transformed to the energy space. We do this by introducing a complex variable $\omega$, which satisfies ${\rm Im}[\omega]>0$. We obtain
\begin{equation}
 C_{1 \rm pt.}(\omega ,T ) 
 \equiv 
 \int_{0}^{T} dt\, e^{i t \omega } C_{1 \rm pt.}(t ) 
= 
i\sum_{n}
\frac{ \braketsub{\ell_f}no{}
\,\braketsub n{\ell_i}{}o
}{
\Delta \omega_n}
\left(1
-e^{i T \, \Delta \omega_n}
\right),
\label{eq:C1pt_omega_cont}
\end{equation}
where $\Delta \omega_n =\omega - E_n$.

In a standard (classical) lattice calculation, time is discretized, therefore, $C_{1 \rm pt.}(t) $ will only be constrained for a finite number of discrete points in time. Assuming there are $N_t$ points evenly spaced by $\Delta t = \frac{T}{N_t}$, and approximating the integral with a sum, we find 
\begin{equation}
 \widetilde{C}_{1 \rm pt.}(\omega , \Delta t,T ) 
 \equiv 
\Delta t \sum_{n_t =1}^{N_t} \, e^{i n_t \Delta t\, \omega } C_{1 \rm pt.}(t ) 
 = 
\Delta t 
\sum_{n}
 \braketsub{\ell_f}no{}
\,\braketsub n{\ell_i}{}o
  \, e^{i  \Delta \omega_n \Delta t} \left(\frac{1-e^{i  \Delta \omega_n T }}{1-e^{i\Delta \omega_n \Delta t}} \right)\ .
\label{eq:C1pt_omega_disc}
\end{equation}
Note that in the limit $\Delta t\to 0$, one recovers Eq.~\eqref{eq:C1pt_omega_cont}. 

\begin{figure}
    \centering
    \includegraphics[width=.8\textwidth]{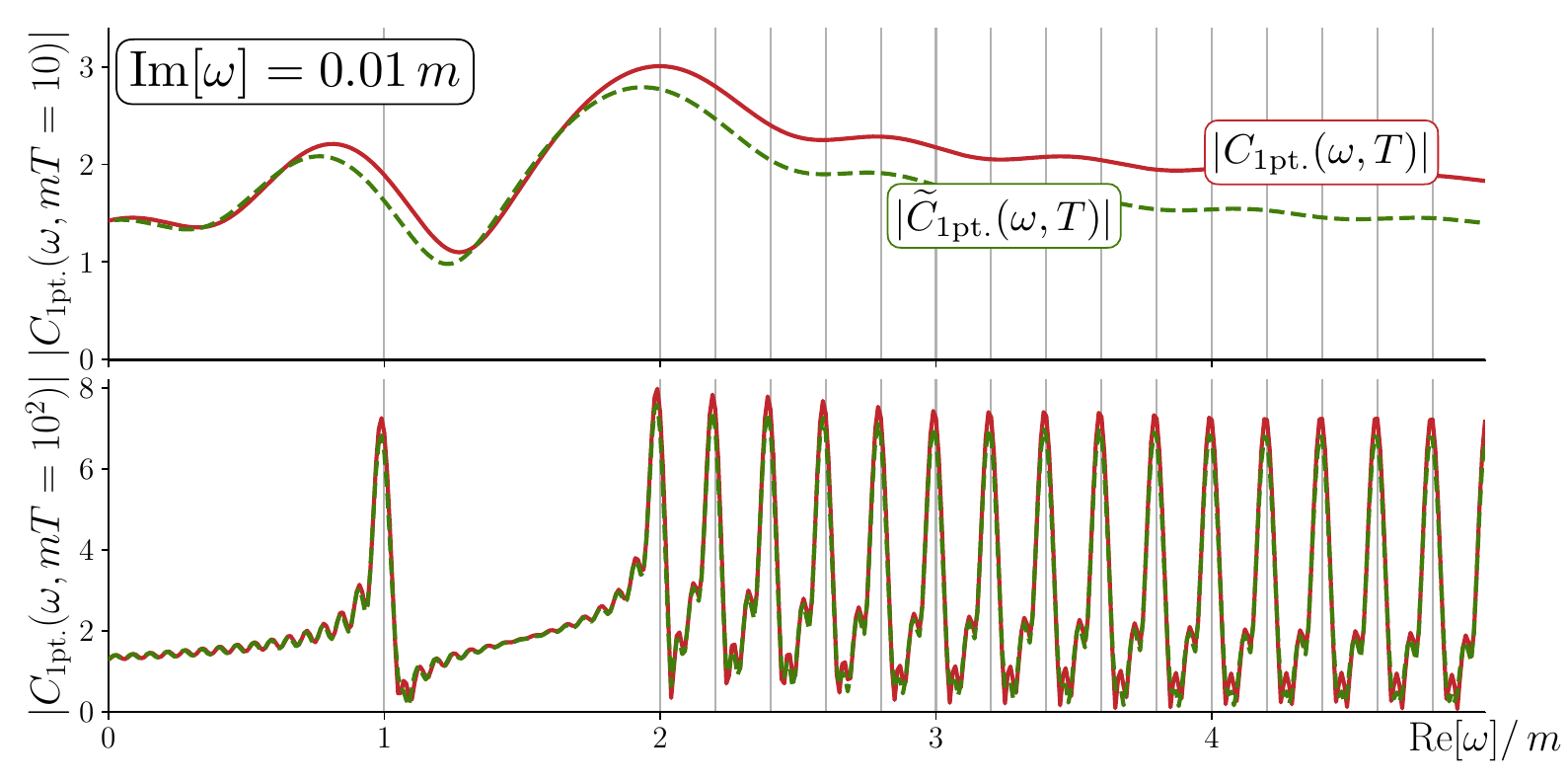}
    \caption{Shown are the values of $C_{1 \rm pt.}(\omega, T)$ [defined in Eq.~\eqref{eq:C1pt_omega_cont}] and  $\widetilde{C}_{1 \rm pt.}(\omega, T)$ [defined in Eq.~\eqref{eq:C1pt_omega_disc}] as red solid lines and green dashed lines respectively. In making these plots, we used a simple toy model with a single stable particle described in the text. The vertical dashed lines are the values of actual spectrum of the model. For $\widetilde{C}_{1 \rm pt.}(\omega, T)$, we used $m\Delta t = 1/10$. The imaginary part of $\omega$ has been fixed to $m/100$. The top panel corresponds to $T =10/m$, while the bottom corresponds to $ T= 100/m$.}
    \label{fig:C1pt}
\end{figure}

An illustrative example is shown in Fig.~\ref{fig:C1pt}. For simplicity, it is assumed $\braketsub{\ell_f}{n}o{}=1/\sqrt{N}$, where $N$ is the number of states considered. The number of states has been truncated to $N=90$, the first few of which are depicted by the vertical grey lines. For simplicity, we have assumed that the incoming free-particle eigenstate has the quantum numbers of a single-particle state, which are different than those of the vacuum of the theory. As a result, the lowest-lying state this will couple is the single-particle state with mass $m$. Assuming a single species in this theory and no deep bound states, the rest of the spectrum would be near the two-particle threshold and/or above it. This means that there is a mass gap of at least $m$ between the ground state and the rest of the states. With this scenario in mind, we let the rest of the states to have energies $\geq 2m$ and assume they are evenly separated with a gap of $m/5$. The solid red lines correspond to $C_{1 \rm pt.}(\omega, T)$ in Eq.~\eqref{eq:C1pt_omega_cont} using ${\rm Im}[\omega]/m=0.01$. The dashed lines correspond to $\widetilde{C}_{1 \rm pt.}(\omega, T)$ in  Eq.~\eqref{eq:C1pt_omega_disc} and $m\Delta t = 1/10$. The results are shown for two values of $T =10/m, 10^2/m$ corresponding to $N_t=10^2, 10^3$. One sees a signal for the ground state, albeit this is distorted by the interference with the excited states. Depending on the mass gap and the resources available allowing for the time resolution, it might be more beneficial to use alternative methods (e.g., the rodeo method~\cite{Choi:2020pdg}) that may provide quicker resolution of the spectrum.

We conclude by remarking that this method provides a systematically improvable way for accessing the spectrum from real-time correlation functions that have been accessed for a finite number of measurements in time. As seen in comparing the top and bottom panels of Fig.~\ref{fig:C1pt}, the ground state resolution improves as $mT$ increases.

\subsection{Accessing local matrix elements currents}\label{secn:I}

This procedure can be generalized to allow for the determination of matrix elements of currents that are local in time. In what follows we assume a generic current, which we label as $\mathcal{J}$ and insert at $t=0$. To specify the initial and final states that couple to the current, we insert two time evolution operators at times $t_i$ and $t_f$ satisfying  $t_f>0>t_i$. Since these correlation functions depend on three-time locations, $(t_i,0,t_f)$, we refer to them as three-point correlators. We can write these explicitly in terms of unitary time-evolution operators and the current,
\begin{align}
C_{3 \rm pts.}(t_f,  t_i) 
&\equiv \brasub{\ell_f}o  e^{-it_f H} \mathcal{J}(0) e^{it_i H}\ketsub{\ell_i}o \ ,
\label{eq:C3pt_def}
\end{align}
making it amenable for quantum simulations using quantum gates.

Inserting a complete set of states, we deduce the spectral decomposition 
\begin{align}
C_{3 \rm pts.}(t_f,  t_i) 
&= \sum_{n_f, n_i}
 e^{-it_f E_{n_f}} 
 e^{it_i E_{n_i}} 
\braketsub{\ell_f}{n_f}o{}\,\braketsub{n_i}{\ell_i}{}0
\bra{n_f}\mathcal{J}(0) \ket{n_i}.
\label{eq:C3pt_spectral}
\end{align}
Working in the time interval $t\in [-T_i,T_f]$, with $T_i,T_f >0$, we obtain the Fourier transform of the three-point correlation function, 
\begin{align}
 C_{3 \rm pts.}(\omega_f,\omega_i, T_f, T_i ) 
 &\equiv 
 \int_{0}^{T_f} dt_f\, e^{i t_f \omega_f }
 \int_{-T_i}^{0} dt_i\, e^{-i t_i \omega_i }
C_{3 \rm pts.}(t_f,  t_i) 
\nn\\
&= \sum_{n_f, n_i}
i\,\frac{\braketsub{\ell_f}{n_f}o{}}{\Delta \omega_{n_f}}
\,i\,\frac{\braketsub{n_i}{\ell_i}{}o }{
\Delta \omega_{n_i}}
\left(1
-e^{i T_f\Delta \omega_{n_f}}
\right)
\left(1
-e^{i T_i\Delta \omega_{n_i}}
\right)
\bra{n_f}\mathcal{J}(0)\ket{n_i}
\label{eq:C3pt_omega}
\end{align}
where we have introduced complex frequencies $\omega_i$ and $\omega_f$, and the frequency differences $\Delta \omega_{n_j} = \omega_j - E_{n_j}$.

By scanning the frequency-dependent correlation function as a function of both $\omega_i$ and $\omega_f$, one can in principle identify the spectrum of the theory since these values of $\omega$ would give the local maxima. More importantly, Eq.~\eqref{eq:C3pt_spectral} tells us how the correlation function at these values of $\omega$ depend on the desired matrix elements of the external current. As $T_f$ and $T_i$ becoming increasingly large, the value of the correlator at a given peak associated with an eigenstate of the theory would be saturated by a single value of $\bra{n_f}\mathcal{J}(0)\ket{n_i}$ and other quantities that may be constrained from one-point correlation functions. 

Here we have assumed that time is a continuous parameter, which in a standard (classical) lattice calculation it is not. Lifting this assumption can be done using the same steps used to arrive at Eq.~\eqref{eq:C1pt_omega_disc}.

\subsection{Accessing non-local current matrix elements}

Finally, we arrive at the matrix elements of current that are displaced in time. In particular, we are interested in accessing matrix elements of the form $\bra{n_f}\mathcal T[\mathcal{J}(t_c)\mathcal{J}(0)]\ket{n_i}$, where two local currents are inserted at different times.

Here we discuss the possibility of accessing these from four-point correlation functions. We will assume a particular time ordering where $t_c > 0$, but this is a soft assumption in what follows that is easily lifted. Assuming $t_f > t_c >0 > t_i$, we can define the desired correlator as
 \begin{align}
C_{4 \rm pts.}(t_f, t_c, t_i) 
&\equiv 
\brasub{\ell_f}o
e^{-i(t_f-t_c) H} \mathcal{J}(0)
e^{-it_c H}
\mathcal{J}(0)
e^{it_i H}
\ketsub{\ell_i}o
\nn\\
&=
\brasub{\ell_f}o
e^{-it_f H} \mathcal{J}(t_c)
\mathcal{J}(0)
e^{it_i H}
\ketsub{\ell_i}o. 
\label{eq:C4pt_def}
\end{align}
The first line provides a representation that is most immediately useful for quantum computations. The second equality follows from the identification of $\mathcal{J}(t_c) = e^{i t_c H} \mathcal{J}(0) e^{-it_cH}$ as a Heisenberg operator. 

By inserting a complete set of states, we find \eq{C4pt_def} can be rewritten, 
\begin{align}
C_{4 \rm pts.}(t_f, t_c, t_i) 
&= \sum_{n_f, n_i}
 e^{-it_f E_{n_f}} 
 e^{it_i E_{n_i}} 
\braketsub{\ell_f}{n_f}o{}\,\braketsub{n_i}{\ell_i}{}o
\bra{n_f}\mathcal{J}(t_c)\mathcal{J}(0)\ket{n_i}.
\label{eq:C4pt_spectral}
\end{align}

At this stage, this looks similar to Eq.~\eqref{eq:C3pt_spectral}, with the only difference being that the matrix element also depends on time. Given our goal is to constrain $\bra{n_f}\mathcal{J}(t_c)\mathcal{J}(0)\ket{n_i}$, we can perform a Fourier transform of this. By determining the correlation function over  continuous time periods given by $t_i\in [-T_i,0]$ and $t_f\in [t_c,T_f]$, one  arrives at 
 \begin{align}
 C_{4 \rm pts.}(\omega_f,\omega_i,t_c, T_f, T_i ) 
 &\equiv 
 \int_{t_c}^{T_f} dt_f\, e^{i t_f \omega_f }
 \int_{-T_i}^{0} dt_i\, e^{-i t_i \omega_i }
C_{4 \rm pts.}(t_f, t_c, t_i) 
\nn\\
&= \sum_{n_f,n_i}
i\,\frac{\braketsub{\ell_f}{n_f}o{}}{\Delta \omega_{n_f}}
\,i\,\frac{\braketsub{n_i}{\ell_i}{}o }{
\Delta \omega_{n_i}}
\left(e^{i t_c\Delta \omega_{n_f}}
-e^{i T_f\Delta \omega_{n_f}}
\right)
\left(1
-e^{i T_i\Delta \omega_{n_i}}
\right) 
\bra{n_f}\mathcal{J}(t_c)\mathcal{J}(0)\ket{n_i}. 
\label{eq:C4pt_omega}
\end{align}
 
 This result closely resembles that of Eq.~\eqref{eq:C3pt_omega}. The key differences are due to the boundary of $t_f$, and the time dependence of the matrix element. Otherwise the key observations made for Eq.~\eqref{eq:C3pt_omega} carry through. In particular, by considering large values of $|T_j|$ and non-zero values of ${\rm Im}[\omega_j]$, this correlation function will have peaks at the spectra of the theory. The value at the peak will depend on the desired matrix elements and quantities that are directly accessible from the one-point correlation function, \eq{C1pt_omega_cont}. In other words, by determining both the one- and four-point correlation functions for a sufficiently large range of time, one would be able to determine real-time long-range matrix elements. 
 
 In arriving at Eq.~\eqref{eq:C4pt_omega}, we have assumed time to be a continous variable. This is an assumption that is simple to lift following the steps taken to go from Eq.~\eqref{eq:C1pt_omega_cont} to Eq.~\eqref{eq:C1pt_omega_disc}.

 \subsection{Comments on the feasibility } 
 The viability of implementing the proposed techniques depends on a range of details on the implementation of the calculations and the underlying dynamics of the theory. On the latter point, investigation such as the one presented in Ref.~\cite{Briceno:2020rar}, where the study of Compton-like amplitudes from real-time correlation functions was first proposed, could provide a guide on the scale of resources needed to be able to carry out studies of realistic theories. In particular, the study considered the implications for strongly-interacting theory in 1+1D theories, and it was found that using spatial volumes satisfying $mL= 20-30$, where $m$ is the lightest mass in the theory, Compton amplitudes may be determined from real-time correlation functions. Using this same framework, one can provide estimates of the values of $mT$ and ${\rm Im}[T \omega_j]$ needed to arrive at percent-level systematic errors. 
 Such an investigation lies outside of the scope of this work, and it might be the case that alternative methods (e.g., rodeo method~ \cite{Choi:2020pdg} or generalizations) might prove to be more beneficial.


\section{Quantum simulation: the case for quantum optics}
\label{secn:2}

\subsection{Introduction} 
\subsubsection{Basic principle} 

The quantum calculation of scattering amplitudes in an interacting relativistic scalar quantum field theory (QFT) was first proposed as a qubit-based quantum sampler~\cite{Jordan2012,Jordan2014}.
That algorithm was subsequently translated to a continuous-variable (CV) sampler~\cite{Marshall2015a}, which paves the way to quantum optical implementations and also allows the direct simulation of one quantum field with only one quantum optical field (rather than incurring the overhead of encoding the $m$ samples of the discretized amplitude of one quantum field into $\log_2 m$ qubits).

The basic idea is to encode a relevant quantum unitary evolution $U(t)$ into a quantum circuit, with specifically prepared input states $\ket{\text{in}}$. The sampling of that quantum circuit, i.e., the measurement of the quantum circuit outcomes $\ket{\text{out}}\bra{\text{out}}$, can be used to reconstruct the correlator
\begin{align}
    \mathcal C(t) = \bra{\text{out}}U(t)\ket{\text{in}}.\label{eq:A}
\end{align}
For concreteness, we will consider a 1+1-dimensional complex QFT with a quartic [$\phi^4$] interaction and a conserved current.  

An essential feature of our approach is that it gives access to phase-sensitive probability amplitudes, \eq A, rather than transition probabilities $\mathcal P(t)=|\mathcal C(t)|^2$ proposed by earlier approaches \cite{Jordan2012,Marshall2015a}. As exposed in Sec.~\ref{secn:obs}, $\mathcal C(t)$ is the most readily interesting quantity because the quantum phase plays a key role in performing the Fourier transform, e.g., \eq{C3pt_omega}, and accessing spectra, among other physical quantities. The problem of determining the phase of the probability amplitudes manifests itself in two ways. First, we need to know the relative phase across different final states, assuming $t$ is fixed. Second, we need to know the relative phase across different values of $t$. This is discussed in \secn{timephase}.

The method discussed below allows for the determination of two relative phases by experimentally interfering with the two corresponding quantum evolutions. This method can be implemented easily for different output states at the same evolution time $t$. For different evolution times, the implementation is more involved and non-Gaussian resources (in terms of their Wigner function) will be needed.

\subsubsection{Quantum advantage} \label{secn:QA}

Before we proceed to the detailed presentation of our quantum simulator, it is important to review the case for using quantum sampling as a quantum simulator of correlators ${C}(t)$  
and probabilities ${P}(t) \propto |C(t)|^2$. This boils down to the question of whether a quantum simulation would yield any quantum advantage. 

As we mentioned earlier, quantum samplers, such as boson samplers~\cite{Aaronson2010} and Gaussian boson samplers~\cite{Hamilton2017}, derive their quantum advantage from their ability to sample from a classically hard-to-calculate probability distribution. Here, we seek to achieve statistically significant sampling in order to reconstruct a probability distribution (and also probability \textit{amplitude} distribution) from the histogram of experimental data, thereby realizing a ``quantum  Galton board.'' The question here is: might the sampling time grow exponentially or worse with the size of the problem, thereby negating any quantum advantage?

Answering this question requires, first, to know how many samples are needed to achieve statistical significance for a given number of distinct sampling outcomes. This question was answered in Ref.~\cite{Weissman2003} where it was shown, in a nutshell, that the probability reconstruction error depends only \textit{linearly} on the total number of sampling outcomes. 

Next, we need to know how this number of sampling outcomes, i.e., of histogram bins, scales with the size of the simulated system. In general, this depends on the measurement basis used to sample and is related to the exponential overhead of quantum state estimation: for $N$ qubits, a single arbitrary state can span all $2^N$ basis vectors, requiring on the order of as many measurements to determine it.

However, whereas a classical computer needs to access the whole $2^N$-dimensional Hilbert space of $N$-qubit quantum states in order to calculate quantum evolution~\cite{Feynman1982}, the quantum simulator of a generally local Hamiltonian~\cite{Lloyd1996} does not, because such Hamiltonians only feature a polynomial number of parameters, leading to ``exploration'' of a polynomial-sized region of Hilbert space only.\footnote{A Hamiltonian with a number of parameters exponential in $N$ simply cannot be efficiently quantum-simulated, of course.} In our case, energy-momentum conservation and the physics behind our QFT system will allow us to make an informed guess of the polynomial-sized region of Hilbert space that our simulator must work in, and therefore we will not be searching blindly. This allows us to expect, in principle, a quantum advantage from the quantum simulations we present here. 

Finally, it is worth reemphasizing the major advantage of quantum simulations over Monte Carlo methods. As will be presented in \secn{timephase}, a crucial result of this work is the phase-sensitive quantum simulation of correlators, such as $C_{1 \text{pt.}}(t)$, Eq.\ \eqref{eq:Ct}. In contrast, standard Monte Carlo classical computing techniques suffer from a sign problem that prohibits a reliable determination of the phase.\footnote{Note that proof-of-principle studies of real-time observables using Monte Carlo techniques have been performed~\cite{Alexandru:2016gsd}.}


\subsection{The physical model}
Turning now to the specific example. We consider a complex massive scalar field $\phi(x) = [\phi_1(x) + i\phi_2(x)]/\sqrt{2}$ of mass $m$ in a single spatial dimension denoted by $x$. We discretize space using $L$ lattice points ($x=0,1,\dots, L-1$), impose periodic boundary conditions, and choose units so that the lattice spacing is $a=1$. 
Let $\pi$ be the conjugate momentum to the complex field $\phi$ obeying the commutation relations
\be [\phi (x) , \pi^\dagger (x') ] = i \delta_{xx'} \ee
The Hamiltonian is
\be H =  \sum_{x=0}^{L-1} \left[  \pi^\dagger (x)\pi (x) + \nabla \phi^\dagger (x) \nabla \phi (x) + m_0^2 \phi^\dagger (x) \phi (x)  + \frac{\lambda}{4} (\phi^\dagger (x) \phi (x) )^2 \right]  \ee
where $\nabla \phi (x) \equiv \phi(x+1) -\phi (x)$, $m_0$ is the bare mass, and $\lambda$ is the coupling constant. To renomalize the theory, $m_0^2$ can in general take negative values. In the procedure that follows, we will diagonalize the quadratic piece of the Hamiltonian exactly. This requires that the mass appears in the Hamiltonian is physical, satisfying $m^2>0$. With this in mind, we introduce the physical mass $m$, and the center terms $\delta m$,
\be \delta_m = m_0^2 - m^2. \ee

We split the Hamiltonian into free and interaction pieces, $H= H_0 + H_\text{int}$, with
\begin{align}
H_0 &=  \sum_{x=0}^{L-1}  \left[ \pi^\dagger (x)\pi (x) + \nabla \phi^\dagger (x) \nabla \phi (x) + m^2 \phi^\dagger (x) \phi (x)  \right] \ ,\label{eq:H0}\\
H_\text{int} &=  \sum_{x=0}^{L-1} \left[  \delta_m \phi^\dagger (x) \phi (x)  + \frac{\lambda}{4} (\phi^\dagger (x) \phi (x) )^2 \right].\label{eq:Hint}
\end{align}
Note, we have introduced the contribution proportional to the counter term into the interaction. Although this is quadratic in the fields, we will treat it differently than the free part of the Hamiltonian.

To diagonalize $H_0$, we write the fields in terms of canonical creation and annihilation operators,
\begin{align}
\phi(x) &= \frac{1}{\sqrt{L}} \sum_{k =0 }^{L-1} \frac{1}{\sqrt{2\omega(k)}} 
\left( b (k) e^{2i\pi k x/L} + c^\dagger (k) e^{-2i\pi k x/L}  \right)  \
,\label{eq:phi}\\ \label{eq:pi}
\pi(x) &= \frac{i}{\sqrt{L}} \sum_{k =0 }^{L-1} \sqrt{\frac{\omega(k)}{2}} \left(
b^\dagger (k) e^{-2i\pi kx/L} - c (k) e^{2i\pi kx/L}\right) \ ,
\end{align}  
and
\be \omega(k) = \sqrt{m^2 + 4 \sin^2 \frac{\pi k}{L}} \label{eq:disp}\ee
is the discretized dispersion relation for a scalar particle. The operator $b^\dagger (k)$ ($c^\dagger (k)$) creates a particle (anti-particle) of physical momentum $p= \frac{2\pi k}{L}$ in the periodic finite volume and energy $\omega (k)$. The  factors of $1/\sqrt{L}$ appearing in front of the sums in \eqs{phi}{pi} were introduced to make the creation and annihilation operators satisfy the  commutation relations
\be [b(k), b^\dagger (k')] = [c(k), c^\dagger (k')] =  \delta_{kk'}. 
\label{eq:bc}\ee
This is in contrast to the more standard $1/L$, which appears in discrete sums of finite-volume momenta. This relative factor of $\sqrt{L}$ is absorbed into the definition of the creation and annihilation operators.

The free Hamiltonian, \eq{H0}, can be diagonalized using these creation and annihilation operators,
\be H_0 = \sum_{k =0 }^{L-1} \omega(k)  \left[ b^\dagger (k) b(k) + c^\dagger (k) c(k) \right] \ , \ee
where we normal-ordered so that the ground state $\ketsub \Omega o$ has vanishing energy. It is annihilated by all annihilation operators ($b(k) \ketsub \Omega o = c(k) \ketsub \Omega o = 0$).
Eigenstates of the free Hamiltonian $H_0$ are constructed by acting on the ground state $\ketsub{\Omega}{o}$ with creation operators. Free particles and anti-particles are created with $b^\dagger (k)$ and $c^\dagger (k)$, respectively \cite{Kubra2018}. For example, states containing $n$ free particles or $\overline n$ free anti-particles each of momentum $\frac{2\pi k}{L}$ are, respectively,

\begin{align}
\label{eq:bk} \ketsub{n}{o,k,b} &= (n!)^{-\frac12} \ b^{\dagger n} (k)\, \ketsub{\Omega}{o} , \\ 
\ketsub{\bar n}{o,k,c} & = (\bar{n}!)^{-\frac12}\ c^{\dagger\bar n} (k)\,\ketsub{\Omega}{o}\ .
\end{align}

\subsection{Quantum simulation using quantum optics}

 The goal of our quantum optical simulator if to evaluate correlators of the form $\brasub{l_f}{o}\exp(-iHt)\ketsub{l_i}{0}$, \eq{Ct}, or current matrix elements, \eq{C3pt_def}, or \eq{C4pt_def}. The latter two contain current operators that will require another layer of sophistication involving non-Hermitian gates. We show how to implement these in \secn{mbqc}, by leveraging the power of measurement-based photonic quantum computing. We start with the unitary evolution of \eq{Ct}.

As \eq{Ct} indicates, we first need to prepare the system in free-field eigenstates, which can be simulated by Fock states of light, as addressed in \secn{lo} below (see \secn{f} for concrete experimental preparation). 

The next step is to implement quantum evolution under the full Hamiltonian $H=H_o+H_\text{int}$. Because $[H_o,H_\text{int}]\neq0$, this will be done using a Trotter-Suzuki expansion in $N$ steps of duration $\Delta t=t/N$ each, implementing the unitary
\be\label{eq:evol} {U} (t) \simeq \left[ U_o(\Delta t) U_\text{int}(\Delta t) \right]^N \equiv \left( e^{-i\Delta t H_0} e^{-i\Delta t H_\text{int}} \right)^N.  \ee
We can now examine the separate implementations $U_o$ and $U_\text{int}$. Implementing $U_o$ is straightforward as we can already place the system in the eigenbasis of $H_o$ by way of the Gaussian transformations in \eqs{pi}{phi}, where Gaussian refers to the Wigner function of the transformation. Implementing $U_\text{int}$ will require a bit more work, especially for the $\phi^4$ Hamiltonian term, and this is where quantum optical simulation can be useful.

We now turn to the quantum optics specifics of our quantum simulation. We start by distinguishing between the quadratic (Gaussian, easy) and the quartic (non-Gaussian, hard) terms of $H_\text{int}$. Any Gaussian Hamiltonian can be implemented by quantum optical squeezing~\cite{Bachor2019}---see also  Appendix~\ref{AppA}---and optical interferometers~\cite{Braunstein2005} or, equivalently, by measurement-based quantum computing (MBQC) over CV cluster states~\cite{Gu2009}. The non-Gaussian Hamiltonian terms can then be accessed by simply adding photon-number-resolving (PNR) measurements to the aforementioned CV MBQC capability, as described in \secn{mbqc}.

\subsubsection{ Optically encoding the free Hamiltonian $H_o$}\label{secn:lo}

Having introduced the QFT of interest, we review the key concepts that allow for the use of quantum optics to diagonalize a massive free scalar QFT, based on techniques discussed in \cite{Marshall2015a}. For a brief introduction to the basics of the quantum optics formalism used here, we point the reader to Appendix~\ref{AppA}. 

The free ground state $\ketsub \Omega o$ of the massive boson field can be straightforwardly simulated by the vacuum state of a photonic system consisting of $2L$ independent optical qumodes. Thus, for our purposes, it is convenient to assign a qumode not to each lattice site (position representation) as was done in Ref.~\cite{Marshall2015a}, but to each site of the dual lattice (momentum representation). We also need to use two qumodes per dual site, corresponding to particles and antiparticles. Thus, we identify, in \eq{vac} below, the free massive boson vacuum (left-hand side) to the photonic qumode vacuum (right-hand side)
\be \label{eq:vac}\ketsub\Omega o \equiv \prod_{k=0}^{L-1} \ketsub{0}{\gamma , k,b} \otimes \ketsub{0}{\gamma , k,c} \ee 
where $(k,i)$ ($i=b,c$) label the optical qumode and $\ketsub 0 \gamma$ is its ground state [Eq.\ \eqref{eq:4} with $n=0$]. Thus, eigenstates of $H_0$ can be constructed as products of photon number eigenstates of the qumodes.
For example, we create the multi-particle states \eqref{eq:bk} with photonic fields as $\ketsub{n}{\gamma,k,b}$ and $\ketsub{\bar n}{\gamma,k,c}$, respectively. Doing this directly requires single-field light emitters of fixed photon numbers $n$ and $\bar n$, which is experimentally arduous, though feasible~\cite{Somaschi2016,Loredo2016}. It is easier to employ photon-pair emitters, such as based on parametric downconversion in nonlinear materials~\cite{Armstrong1962}. Such sources can emit distinguishable fields, thereby creating photon-number correlated (and field-entangled) qumodes, whose quantum state is the two-mode-squeezed state described in \eq{tms} of the Appendix. Because of the photon number correlation, measuring the photon number in one qumode will project the other qumode onto a Fock state, thereby generating a fixed number of simulated particles or antiparticles. In \secn{mbqc}, we will detail a specific implementation of this protocol that leverages measurement-based quantum computing over cluster entangled states. 

\subsubsection{ Optically encoding the interaction Hamiltonian $H_\text{int}$}

We now turn to the implementation of the interaction Hamiltonian $H_{\text{int}}$, \eq{Hint}. The quartic $\phi^4$ interaction, which requires non-Gaussian gates, makes this a nontrivial task. Expressing $H_{\text{int}}$ in terms of the modes of the free Hamiltonian results in a rather complicated expression involving terms that mix different qumodes. To implement quantum gates involving $H_{\text{int}}$, we introduce a Gaussian unitary transformation $G$ that reduces the complexity of $H_{\text{int}}$ allowing its efficient implementation with single and two-qumode gates. 

The form of the interaction Hamiltonian, \eq{Hint}, strongly suggests 
to define creation and annihilation operators at each lattice site (position representation) instead of on the dual lattice (momentum representation), as was used to diagonalize the free Hamiltonian. We will therefore need to relate the corresponding qumodes, which we show can be done easily with quantum optics.

At each lattice site $x$, we introduce quadratures for two qumodes as follows. We define their annihilation operators,
\begin{equation}
\label{eq:30} 
B (x) = \frac{\phi(x) +  i \pi^\dag (x) }{\sqrt{2}} \ ,\ 
\
C (x) = \frac{\phi^\dag(x) + i\pi (x)}{\sqrt{2}} \ ,
\end{equation}
which obey standard Heisenberg algebras, $[B(x), B^\dag (x')] = [C(x) , C^\dag (x') ] = \delta_{xx'}$. The quadratures are
\begin{equation}\label{eq:QPbc}
    Q_{B} (x) = \frac{B(x) + B^\dag (x) }{\sqrt{2}} \ , \ \ P_{B} (x) = \frac{B(x) - B^\dag (x)}{\sqrt{2} i} \ , \ \ Q_{C} (x) = \frac{C(x) + C^\dag (x) }{\sqrt{2}} \ , \ \ P_{C} (x) = \frac{C(x) - C^\dag (x)}{\sqrt{2} i} \ ,
\end{equation}
in terms of which the interaction Hamiltonian \eqref{eq:Hint} can be written as
\be\label{eq:Hint2} H_{\text{int}} = \sum_{x=0}^{L-1} \left( \frac{\delta_m}{2} \left\{[Q_B(x)+Q_C(x)]^2 + [P_B(x) - P_C(x)]^2\right\} + \frac{\lambda}{16} \left\{[Q_B(x)+Q_C(x)]^2 + [P_B(x) - P_C(x)]^2\right\}^2 \right) \ee
Notice that $Q_B(x)+Q_C(x)$ and $P_B(x) - P_C(x)$ commute~\cite{Bohr1935}, so ordering is not important.
To implement the various terms in \eq{Hint2}, we need to engineer the quadratures $(Q_{B,C}, P_{B,C})$ from our original photonic fields which, again, encode the linear momenta of the free field, \eqlist{phi}{bc}. To construct the transformation relating the two sets of quadratures, we start by relating the corresponding annihilation operators.
Using \eqs{phi}{pi}, it is straightforward to obtain
\begin{align}
B(x) &= \frac{1}{2\sqrt{L}} \sum_{k=0}^{L-1} \left[ \left( \frac{1}{\sqrt{\omega (k)}} + \sqrt{\omega (k)} \right) b (k) + \left( \frac{1}{\sqrt{\omega (k)}} - \sqrt{\omega (k)} \right) c^\dagger (L-k) \right] e^{2i\pi kx/L} \nonumber\\  
C(x) &= \frac{1}{2\sqrt{L}} \sum_{k=0}^{L-1} \left[ \left( \frac{1}{\sqrt{\omega (k)}} - \sqrt{\omega (k)} \right) b^\dag (k) + \left( \frac{1}{\sqrt{\omega (k)}} + \sqrt{\omega (k)} \right) c (L-k) \right] e^{2i\pi kx/L}  
\end{align}
With these expressions, we can immediately read off the Fourier transforms of $B(x)$ and $C(x)$ ,
\begin{align}\label{eq:37}
\widetilde{B} (k) &= \frac12\left( \frac{1}{\sqrt{\omega (k)}} + \sqrt{\omega (k)} \right) b (k) + \frac12\left( \frac{1}{\sqrt{\omega (k)}} - \sqrt{\omega (k)} \right) c^\dagger (L-k) \nonumber\\
\widetilde{C} (k) &= \frac12\left( \frac{1}{\sqrt{\omega (k)}} - \sqrt{\omega (k)} \right) b^\dag (k) + \frac12\left( \frac{1}{\sqrt{\omega (k)}} + \sqrt{\omega (k)} \right) c (L-k).
\end{align}
Note that the discrete Fourier transform can be straightforwardly implemented as a passive optical unitary transformation $U$  
over $2L$ modes, realized by a $2L\times2L$ generalized optical interferometer~\cite{Clements2016}. Thus,
\be\label{eq:38} U^\dag \bm{B} U = \widetilde{\bm{B}} \ , \ \ U^\dag \bm{C} U = \widetilde{\bm{C}}\ , \ee 
where $\bm{B} = [B(0), B(1), \dots , B(L-1)]^T$, and similarly for $\bm{C}$, as well as $\widetilde{\bm{B}}$ and $\widetilde{\bm{C}}$. Note that $U$ is an operator acting on each component of $\bm{B}$ and $\bm{C}$ individually and is implemented with qumode gates.

Going back to the analytic expressions \eqref{eq:37}, these bear strong significance in quantum optics for they constitute the exact Heisenberg-picture Bogoliubov transformation effected by the two-mode squeezing process (by stimulated emission of pairs of distinguishable photons, see Appendix \ref{AppA}). Using \eq{bogo1}, we thus identify
\begin{equation}
\cosh [r(k)] \equiv\frac12\left(\frac{1}{\sqrt{\omega (k)}} + \sqrt{\omega (k)}\right)\ ,\ \
\sinh [r(k)] \equiv\frac12\left(\frac{1}{\sqrt{\omega (k)}} - \sqrt{\omega (k)}\right)\ , \ \ 
r(k) = -\frac{1}{2} \ln \omega (k), \label{eq:LTUaE}
\end{equation}
where $r(k)$ is the effective squeezing parameter.

This relation between the squeezing parameter and the dispersion relation \eq{disp} illuminates the physical significance of quantum optical squeezing in the context of quantum simulation of massive bosons: on one hand, \textit{optical squeezing confers effective mass to the massless simulating photon field}, on the other hand, \textit{two-mode squeezing generates number correlations between particles of momentum $k$ and antiparticles of momentum $L-k$, $\forall k$}. Formally, this can also be expressed in matrix representation
\begin{equation}\label{eq:39}
\begin{pmatrix}\widetilde{B} (k) \\ \widetilde{C}^\dag (k)\end{pmatrix}
= S^\dagger (k, L-k;r(k)) \
\begin{pmatrix} b(k)  \\  c^\dag(L-k) \end{pmatrix}
\ S(k,L-k;r(k)) =
\begin{pmatrix}\cosh [r(k)]&\sinh [r(k)]\\ \sinh [r(k)]&\cosh [r(k)]\end{pmatrix}
\begin{pmatrix}b (k)\\c^\dagger (L-k)\end{pmatrix}.
\end{equation}
where $S$ is a two-mode squeezing operator (for details, see Appendix \ref{AppA}).

Combining with the discrete Fourier transform, \eq{37}, the initial Fock-state qumodes, which co\"incide with the eigenmodes of the free Hamiltonian, can be transformed into qumodes that simplify the interaction Hamiltonian by the Gaussian transformation $G$,
\begin{equation}
    \bm{B} = G^\dag\ \bm{b}\ G \ , \ \ C = G^\dag\ \bm{c}\ G \ , \ \ G = S\ U\ , \label{eq:G}
\end{equation}
where $G$ acts on individual components of $\bm{b} = [b(0),b(1),\dots, b(L-1)]^T$ and $\bm{c} = [c(0),c(1),\dots, c(L-1)]^T$. Note that any $N$-mode Gaussian transformation in quantum optics can be experimentally realized with $N\times N$ interferometers and single-mode squeezers, according to the Bloch-Messiah singular-value decomposition~\cite{Braunstein2005}. Another way, as we will see, is to use the machinery of measurement-based quantum computing over qumode cluster states. 

The Gaussian transformation $G$ can also be used to relate the quadratures \eqref{eq:QPbc} to the quadratures of the modes in the momentum representation, $b(k), c(k)$. We obtain
\begin{equation}\label{eq:Ga}
    Q_B = G^\dag q_b G \ , \ \ P_B = G^\dag p_b G \ , \ \ Q_C = G^\dag q_c G \ , \ \ P_C = G^\dag p_c G
\end{equation}
where $q_b(k) = [b(k)+b^\dag (k)]/\sqrt{2}$, $p_b(k) = i[b^\dag (k)-b(k)]/\sqrt{2}$, and similarly for $(q_c,p_c)$. This transformation allows us to implement gates involving the quadratures $(Q_{B,C}, P_{B,C})$ using the gates with quadratures $(q_{b,c}, p_{b,c})$ that are native to our photonic substrate. In particular, the interaction Hamiltonian, \eq{Hint2}, can be written in terms of these quadratures as
\be\label{eq:Hint3} H_{\text{int}} = G^\dag \ \sum_{k=0}^{L-1}  h_{\text{int}} (q_b(k), p_b(k), q_c(k) , p_c(k)) \ G \ee 
where
\be\label{eq:Hint4} h_{\text{int}} (q_b, p_b, q_c , p_c) = \frac{\delta_m}{2} [(q_b +q_c)^2 + (p_b - p_c)^2] + \frac{\lambda}{16} [(q_b +q_c)^2 + (p_b - p_c)^2]^2  \ee
The Gaussian unitary $G$ and quantum gates involving various terms of $h_{\text{int}}$ can be implemented with optical elements. Unlike the free Hamiltonian, $h_{\text{int}}$ is not diagonal. However, it is a simplified expression compared to $H_{\text{int}}$ as terms labeled by different values of the index $k$ are now separated and commute with each other. Thus, the contributions for each $k$ can be implemented with quantum gates involving single qumodes and beam splitters. 

For the sake of giving the simplest illustrative example, we now give the equivalent matrix representation of this transformation for two lattice sites.

\subsubsection{Example of two lattice sites}

As an example, consider the case of two lattice sites ($L=2$). We have four modes, labeled $(k,b)$ for particles and $(k,c)$ for anti-particles with $k=0,1$ (stationary and moving (anti-)particle, respectively). We will simulate them with four photonic qumodes. Switching to position representation, we consider modes that are local to lattice sites, $B(x), C(x)$ ($x=0,1$), \eq{30} for $L=2$, that can be simulated with photonic fields by relating them to modes $b(k), c(k)$ ($k=0,1$) via \eq{G} for $L=2$.
Defining $\bm{B} = (B(0),B(1))$, $\bm{C} = (C(0),C(1))$, $\widetilde{\bm{B}} = (\widetilde{B}(0),\widetilde{B}(1))$, $\widetilde{\bm{C}} = (\widetilde{C}(0), \widetilde{C}(1))$, $\bm{b} = (b(0),b(1))$, and $\bm{c} = (c(0), c(1))$, the discrete Fourier transform amounts to
\be B(x) = \frac{1}{\sqrt{2}} ( \widetilde{B} (0) + (-)^x \widetilde{B} (1) ) \ ,\ \ C(x) = \frac{1}{\sqrt{2}} ( \widetilde{C} (0) + (-)^x \widetilde{C} (1) ) \ee
and can be implemented with a pair of 50:50 beam splitters acting on modes $\{ (0,b), (1,b)\}$ and $\{ (0,c), (1,c)\}$, respectively.

The squeezing, \eq{39}, is implemented with two two-mode squeezer, $S_2(r(k))$.  The first two-mode squeezer, $S_2(r(0))$, would be for modes $\{ (0,b), (1,c) \}$ with squeezing parameter $r(0) = - \frac{1}{2} \ln m$, and the other, $S_2(r(1))$, for modes $\{ (1,b), (0,c) \}$ with squeezing parameter $r(1) = - \frac{1}{4} \ln (4+ m^2)$.

\begin{figure}[t]
    \centering
\[\Qcircuit @C=2.8em @R=2em { \lstick{(0,b)} & \qw &\qw & \multigate{1}{S_2 (r(0))}  &\qw & \qw & \multigate{1}{BS} &   \qw  &\qw & \rstick{B(0)} \\  \lstick{\ (1,b)} & \qw & \link{1}{-1} & \ghost{S_2 (r(0))} & \qw & \link{1}{-1} & \ghost{BS} &  \qw  &\qw & \rstick{B(1)} \\ \lstick{(1,c)} & \qw & \link{-1}{-1} & \multigate{1}{S_2 (r(1))} & \qw & \link{-1}{-1} & \multigate{1}{BS} & \qw  &\qw & \rstick{C(1)} \\ \lstick{(0,c)} & \qw  & \qw  & \ghost{S_2 (r(1))} & \qw & \qw & \ghost{BS}   & \qw   &\qw & \rstick{C(0)} } \]
    \caption{Quantum circuit (left to right) implementing the unitary $G$, \eq G, for two lattice sites. The notation for the states are defined in the text, $S_2(r(0))$ and $S_2(r(1))$ are two-mode squeezers with squeezing parameters $r(0) = - \frac{1}{2} \ln  m$ and $r(1) = - \frac{1}{4} \ln (4+ m^2)$, respectively, and $BS$ is a 50:50 beam splitter.   } 
    \label{fig:qc1}
\end{figure}
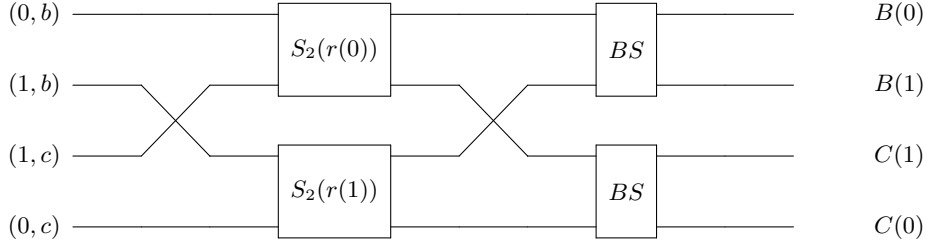

The circuit for the Gaussian transformation, \eq{G} for $L=2$, is shown in \fig{qc1}.

\subsubsection{Quantum computation of correlators}

Finally, we turn to the quantum computation of the correlation functions of the form $\brasub{l_f}{o}\exp(-iHt)\ketsub{l_i}{o}$, \eq{Ct}, discussed in Section \ref{secn:obs}, using photonic qumodes. Having constructed eigenstates of the free Hamiltonian $H_0$ (\secn{lo}), we need to evolve them with the full Hamiltonian $H$. Note that current correlators, such as \eq{C4pt_def}, require another layer of sophistication that requires non-Gaussian (Hermitian) gates. We show how to implement these in \secn{mbqc}, by leveraging the power of measurement-based photonic quantum computing.

 The full Hamiltonian evolution will be done using the Trotter-Suzuki expansion of \eq{evol}. An important question here is what the minimal required sampling time $\Delta t$ is and whether it is accessible to an optical simulator. One reasonable estimate is  $\Delta t\ll \max[\omega(k)]^{-1}\sim 1/m$. Using \eq{LTUaE}, we get $\Delta t\ll e^{-2r(0)}$. 
(Note that we are working in units in which the lattice spacing is $a=1$, so all quantities are dimensionless.)

It is straightforward to implement the evolution unitary $e^{-i\Delta t H_0}$, because $H_0$ is diagonal in the photonic qumodes. This is thus free field evolution, i.e., light propagation, simply defined by the optical path. It can be implemented using phase shifts (a.k.a.\ phase-space rotations), 
\be\label{eq:33} R_b(k;\theta) = e^{i\theta b^\dagger (k) b(k)} \ , \ \ R_c(k;\theta) = e^{i\theta c^\dagger (k) c(k)} \ee acting on particle and anti-particle modes, respectively, as
\be e^{-i\Delta t H_0} = 
\bigotimes_{k=0 }^{L-1}
\, R_b(k; \omega(k)\Delta t) \otimes R_c(k; \omega(k)\Delta t). \ee

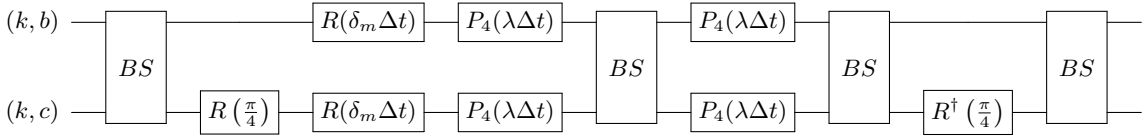
\begin{figure}[t]
    \centering
\[\Qcircuit @C=1.4em @R=2em { \lstick{(k,b)}   & \multigate{1}{BS}  &\qw & \gate{R(\delta_m\Delta t)} & \gate{P_4 (\lambda \Delta t)} & \multigate{1}{BS} & \gate{P_4 (\lambda \Delta t)} & \multigate{1}{BS} & \qw & \multigate{1}{BS}  &\qw  \\  \lstick{\ (k,c)}   & \ghost{BS} & \gate{R \left(\frac{\pi}{4}\right) } & \gate{R(\delta_m\Delta t)} & \gate{P_4 (\lambda \Delta t)} & \ghost{BS} & \gate{P_4 (\lambda \Delta t)} & \ghost{BS} & \gate{R^\dag \left(\frac{\pi}{4}\right) } & \ghost{BS}   &\qw   } \]
    \caption{Quantum circuit implementing the unitary $P_{\text{int}}$, \eq{Pint}, on two qumodes labeled $(k,b)$ and $(k,c)$. BS is a 50:50 beam splitter, $R$ is phase-space rotation given in Eq.~\eqref{eq:33}, and $P_4$ is a quartic phase gate (Eq.~\eqref{eq:P4}). } 
    \label{fig:Pint}
\end{figure}

To simulate $H_\text{int}$ with photonic qumodes, we need the Gaussian unitary $G$, \eq G. Using \eqs{Hint3}{Hint4}, we obtain
\be\label{eq:Pint} e^{-i\Delta t H_\text{int}} 
= G^\dag\    \bigotimes_{k=0 }^{L-1} \, P_{\text{int}} (k) \ G \ , \ \ P_{\text{int}} = e^{-i\Delta t\, h_{\text{int}} (q_b,p_b, q_c,p_c)}    \ee
To implement $P_{\text{int}}$, we massage $h_{\text{int}}$ with these steps:
\begin{align}
    h_{\text{int}} &= BS\ \left[ {\delta_m} (q_b^2 + p_c^2) + \frac{\lambda}{4} (q_b^2 +  p_c^2)^2 \right] \ BS \nonumber\\
    &= BS\ R_c^\dag \left(\frac{\pi}{4}\right) \left[ {\delta_m} (q_b^2 + q_c^2) + \frac{\lambda}{4} (q_b^2 +  q_c^2)^2 \right] \ R_c \left(\frac{\pi}{4}\right)\ BS \nonumber\\
    &= BS\ R_c^\dag \left(\frac{\pi}{4}\right) \left[ {\delta_m} (q_b^2 + q_c^2) + \frac{\lambda}{24} \left( 4q_b^4 +  4q_c^4 + (q_b+q_c)^4 + (q_b-q_c)^4 \right) \right] \ R_c \left(\frac{\pi}{4}\right)\ BS
\end{align}
where $BS$ is a 50:50 beam splitter, and $R_c \left(\frac{\pi}{4}\right)$ is the phase shift defined in \eq{33} with $\theta = \frac{\pi}{4}$. It follows that $P_{\text{int}}$ can be implemented with beam splitters and rotation gates, which are Gaussian, as well as non-Gaussian quartic phase gates,
\be P_4 (\gamma) = e^{i\frac{\gamma}{6} q^4}. 
\label{eq:P4}
\ee 
The circuit implementing $P_{\text{int}}$ is shown in \fig{Pint}.

Cubic phase states and gates can be made by combining two-mode squeezing and PNR detection, as proposed in Ref.~\cite{Gottesman2001}. Optical cubic gates have long been deemed difficult to implement, due to the necessity of detecting at least 50 photons for high enough non-Gaussian quality (measured as negativity of the Wigner function)~\cite{Ghose2007}, but the recent demonstration of PNR detection of up to 100 photons~\cite{Eaton2023} has opened up new possibilities.  Higher-order gates can then be obtained by use of the Baker-Campbell-Hausdorff formula~\cite{Lloyd1999}. It has been shown that 15 cubic phase gates are needed to exactly create one quartic phase gate~\cite{Kalajdzievski2019}. There are, however, strong reasons to believe that other, approximate techniques could be used to lead to a greater economy of resources: for example, approximate probabilistic methods for cubic phase gates have been proposed~\cite{Yukawa2013,Marshall2015}. There has also been a deterministic non-Gaussian gate based on a  tailored Kerr nonlinearity proposed in Ref.~\cite{PhysRevLett.124.240503}.   

\begin{figure}[ht!]
\[\Qcircuit @C=1.0em @R=1em 
{\lstick{\ketsub{m_0}{0,b}}
& \qw & \multigate{4}{G} &\multigate{1}{P_{\text{int}} (0)}  & \multigate{4}{G^\dagger} &\qw & \gate{R(\omega(0)\Delta t)} & \qw & \push{\dots\quad} &  \qw &\measureD{n_0} \\ \lstick{\ketsub{\overline m_0}{0,c}} 
& \qw & \ghost{G} &\ghost{P_{\text{int}} (0)}  & \ghost{G^\dagger} &\qw & \gate{R(\omega(0)\Delta t)} & \qw & \push{\dots\quad} &  \qw &\measureD{\overline n_0} \\ 
\push{\vdots} &\qwred[0] & \qwred[0] &\push{\vdots}  &\qwred[0] &\qwred[0] & \push{\vdots} & \qwred[0] & \qwred[0] & \qwred[0] & \push{\vdots} \\ 
\lstick{\ketsub{m_{L-1}}{L-1,b}} 
& \qw & \ghost{G} &\multigate{1}{P_{\text{int}} (L-1)}  & \ghost{G^\dagger} &\qw & \gate{R(\omega(L-1)\Delta t)} & \qw & \push{\dots\quad} & \qw & \measureD{n_{L-1}}\\ 
\lstick{\ketsub{\overline m_{L-1}}{L-1,c}} 
& \qw & \ghost{G} &\ghost{P_{\text{int}} (L-1)}  & \ghost{G^\dagger} &\qw & \gate{R(\omega(L-1)\Delta t)} & \qw & \push{\dots\quad} & \qw   
& \measureD{\overline n_{L-1}} 
}
\]
    \caption{Quantum circuit displaying one time step $\Delta t$ of the Trotterized evolution operator of \eq{evol}. Each horizontal wire is an optical qumode parametrized as $(k,i)$ ($k=0,\dots,L-1$ and $i=b,c$) initialized on the left in a Fock state. Gates labeled $G$ (\eq{G}), $R$ (\eq{33})  are Gaussian; $P_{\text{int}}$ (\eq{Pint}) involves non-Gaussian quartic phase gates (see \fig{Pint}). To the right are PNR measurements for each qumode.}
    \label{fig:qc2}
\end{figure}
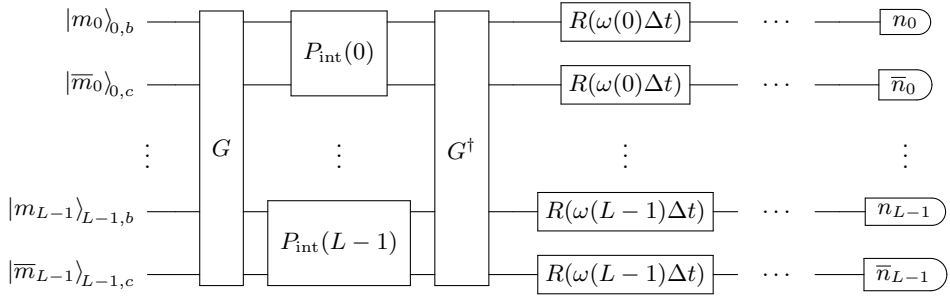

An optical-circuit implementation of the field-encoded evolution operator \eqref{eq:evol} is shown in \fig{qc2}. Note that, though we detail an actual physical optical implementation in \fig{qc2} for the sake of  concreteness, our optical approach will substantially differ, being measurement-based rather than circuit-based quantum computing and will be detailed in \secn{mbqc}. The reason why we propose this measurement-based approach is that it allows the implementation of difficult gates, i.e., the quartic phase gate and current operators, with greater ease and in a near-deterministic manner. 

\section{Coherent quantum simulation of correlators}\label{secn:timephase}

This section details a main result of this paper, which is the method for the quantum simulation of \textit{phase-sensitive} correlators, such as 
$C_{1 \rm pt.}(t) =\brasub{\ell_{f}}{o} e^{-itH} \ketsub{\ell_{i}}{o}$ in  \eq{Ct}, which enables the computation of the Fourier transform in \eq{C1pt_omega_cont}. This requires an additional layer of complexity as, again, quantum sampling alone can only give access to the probability $\mathcal P(t)=|C(t)|^2$ corresponding to a correlator ${C} (t)$. Unsurprisingly, this layer involves quantum interferometry. We will first determine $C(t)$ at a given time $t$ but various initial and final states, and then use these results to determining $C(t)$ at different times.

We can gain access to $C_{1 \rm pt.}(t)$ by performing a series of weak displacements of the state before counting photons. 
The initial and final states in the correlator \eqref{eq:Ct} are eigenstates of the free Hamiltonian $H_0$, and can be written as tensor products of $2L$ Fock states, each simulated by a separate photonic qumode  \eqref{eq:4},
\be\label{eq:59} \ketsub {l_i} o = \bigotimes_{k=0}^{L-1}  \ketsub {m_k} {\gamma, k, b} \otimes \ketsub {\overline{m}_k} {\gamma, k, c} \ , \ \ \ketsub {l_f} o = \bigotimes_{k=0}^{L-1} \ketsub {n_k} {\gamma, k, b} \otimes \ketsub {\overline{n}_k} {\gamma, k, c} \ee
with the initial state containing $m_k$ particles and $\overline{m}_k$ anti-particles with momentum $p  = 2\pi k/L$, and similarly for the final state.
It is convenient to introduce the vectors $\bm{m} = (m_0,\dots, m_{L-1}, \overline{m}_0 , \dots, \overline{m}_{L-1})^T$, $\bm{n} = (n_0,\dots, n_{L-1}, \overline{n}_0 , \dots, \overline{n}_{L-1})^T$, and identify
\be\label{eq:66a} \ketsub{\ell_i}{o} = \ketsub{\bm{m}}{\gamma} \ , \ \ \ketsub{\ell_f}{o} = \ketsub{\bm{n}}{\gamma} \ . \ee

\subsection{Single-mode implementation}

In order to capture the essence of our approach without cluttering the notation, we first consider a toy model with a single physical mode (unlabeled here to simplify notations) and correlator
\begin{equation}\label{eq:60}
    {C}_{n,m} (t)
=\brasub{n}\gamma  e^{-itH} \ketsub{m}\gamma. 
\end{equation}
As described in \secn{mbqc}, we can prepare many copies of $\ket{m}_\gamma$ for a given $m$, evolve the state via MBQC on a 1D chain cluster state, perform PNR detection, and histogram the results against the output photon number $n$, thereby obtaining an empirical probability distribution for ${P}_{n,m}=|{C}_{n,m}|^2$ for each $n$. This can be done efficiently if the number of different outcomes $n$ stays low, in particular when the number of modes increases. But this does not give us access to the full complex number that is ${C}_{n,m}$. To remedy this, we introduce field displacements of the evolved photon state just before measurement,
\begin{equation}
    D(\xi)=e^{\xi {a}^\dag - \xi^*{a}},
\end{equation}
where $|\xi|$ is the amplitude of the added laser field and $\arg(\xi)$ is its phase. 
Field displacements can be straightforwardly implemented experimentally using an unbalanced beamsplitter to coherently add a phase- and amplitude-controlled laser field to a qumode, thereby displacing it in phase space. 
Thus, we are led to consider the modified correlators and corresponding detection probabilities that we have access to,
\be \mathcal{C}_{n,m} (\xi; t) = \brasub{n}\gamma   D(\xi) e^{-itH} \ketsub{m}\gamma \ , \ \ \mathcal{P}_{n,m} = |\mathcal{C}_{n,m}|^2 \ . \ee
For small displacements, $|\xi|\ll1$, we 
can approximate the displacement operator to first order 
\begin{equation}\label{eq:smaldisp}
    D(\xi)\simeq\mymathbb{1}+\xi a^\dag - \xi^* a + \mathcal{O}(\xi^2).
\end{equation} 
and the detection probabilities can be expressed in terms of the correlator \eqref{eq:60} that we are interested in,
\begin{equation}
\mathcal{P}_{n,m}(\xi)= |{C}_{n,m}|^2+2\sqrt{n}\,\text{Re}[\xi^\ast {C}_{n,m} {C}^\ast_{{n-1},m}]-2\sqrt{n+1}\,\text{Re}[\xi {C}_{n,m} {C}^\ast_{{n+1},m}]+\mathcal{O}(\xi^2)\ .
\end{equation}
Suppose $\xi \in \mathbb{R}$. With just three measurement settings, we obtain
\begin{align}
    \mathcal{P}_{n,m}(0)&=| {C}_{n,m}|^2\nonumber\\
    \mathcal{P}_{n,m}(\xi)&=\mathcal{P}_{n,m}(0)+2\xi(\sqrt{n}\,\text{Re}[ {C}_{n,m} {C}^\ast_{{n-1},m}]-\sqrt{n+1}\,\text{Re}[ {C}_{n,m} {C}^\ast_{{n+1},m}])\nonumber\\
    \mathcal{P}_{n,m}(i\xi) &= \mathcal{P}_{n,m}(0) +2\xi(-\sqrt{n}\,\text{Im}[ {C}_{n,m} {C}^\ast_{{n-1},m}]+\sqrt{n+1}\,\text{Im}[ {C}_{n,m} {C}^\ast_{{n+1},m}]).\label{eq:Pnmixi}
\end{align}
Starting with zero-photon ($n=0$) detection events, we obtain a system coupling only two correlators,
\begin{align}
    \mathcal{P}_{0,m}(0)&=| {C}_{0,m}|^2\nonumber\\
    \mathcal{P}_{0,m}(\xi)&=\mathcal{P}_{0,m}(0) -2\xi\,\text{Re}[ {C}_{0,m} {C}^\ast_{{1},m}]\nonumber\\
    \mathcal{P}_{0,m}(i\xi)&=\mathcal{P}_{0,m}(0)+2\xi\,\text{Im}[ {C}_{0,m} {C}^\ast_{{1},m}]\ ,
\end{align}
which is sufficient to determine the correlators ${C}_{0,m}$ and ${C}_{1,m}$ up to a global phase. Substituting these results into the system \eqref{eq:Pnmixi} for $n=1$ yields ${C}_{2,m}\in\mathbb C$, and so on. For a maximum photon detection of $M$ photons, these $3M$ equations can be solved to obtain all ${C}_{n,m}$ correlators involving fewer than $M$ photons.

\subsection{Multi-mode extension}

\begin{figure}[t]
    \centering
    \[\Qcircuit @C=2.8em @R=1em { \lstick{\ketsub{m_0}{\gamma,0,b}} & \qw  & \multigate{3}{e^{-itH}}  &\qw  & \gate{D(\xi_0)} &   \qw   & \measureD{n_0} \\  \lstick{\ketsub{m_1}{\gamma,1,b}} & \qw  & \ghost{e^{-itH}} & \qw  & \gate{D(\xi_1)} &  \qw   & \measureD{n_1} \\ \lstick{\ketsub{\overline{m}_0}{\gamma,0,c}} & \qw  & \ghost{e^{-itH}} & \qw  & \gate{D(\overline{\xi}_0)} & \qw   & \measureD{\overline{n}_0} \\ \lstick{\ketsub{\overline{m}_1}{\gamma,1,c}}  & \qw  & \ghost{e^{-itH}} & \qw  & \gate{D(\overline{\xi}_1)}   & \qw    & \measureD{\overline{n}_1} } \]
    \caption{Phase-sensitive simulation of the one-point correlator \eqref{eq:Ct} for two lattice sites requiring four qumodes.}
    \label{fig:disp}
\end{figure}
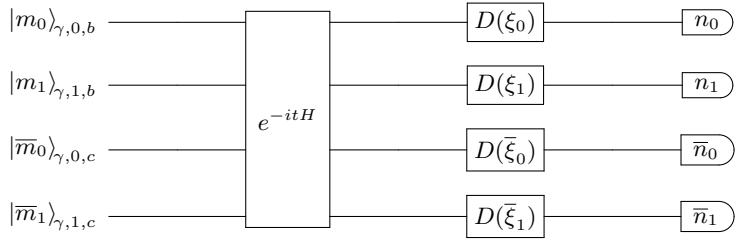

The above procedure can be easily generalized to the full $2L$-mode system, where local weak displacements of $0, \xi,$ and $i\xi$ are applied in every combination to each mode before detection. In this case, the modified correlators and corresponding multi-mode detection probabilities can be written as
\begin{equation}
 \mathcal{C}_{\bm{n},\bm{m}} (\bm{\xi} ; t) = \brasub{\bm{n}}{\gamma} \prod_{k=0}^{L-1} \otimes D_{k,b}(\xi_k)\otimes D_{k,c} (\overline{\xi}_k)   \cdot e^{-itH} \ketsub{\bm{m}}{\gamma} \ , \ \ \mathcal{P}_{\bm{n},\bm{m}}= \left| \mathcal{C}_{\bm{n},\bm{m}} (t) \right|^2,
\end{equation}
where $\bm{\xi} = (\xi_0, \dots, \xi_{L-1} , \overline{\xi}_0 , \dots , \overline{\xi}_{L-1} ) \in \mathbb{C}^{2L}$ is the vector defining the $2L$-mode displacement. 
The PNR detectors project the state of the system onto the $2L$-mode Fock state of the photon field $\ketsub{\bm{n}}{\gamma}$ (Eq.\ \eqref{eq:66a}), and $\ketsub{\bm{m}}\gamma$ is the initial multi-mode photon-number eigenstate before evolution. The quantum circuit for two lattice sites requiring four qumodes is depicted in Fig.\ \ref{fig:disp}.

If only one mode is displaced at a time, we can consider displacements by $0, \xi_j, i\xi_j$, for a given mode $j$. Working as with the single-mode toy model, we obtain expressions of detection probabilities in terms of the one-point correlators \eqref{eq:Ct} that we are interested in,
\begin{align}\label{eq:75}
    \mathcal{P}_{\bm{n},\bm{m}}(\bm{0}) &=| {C}_{\bm{n},\bm{m}}|^2 \nonumber\\
    \mathcal{P}_{\bm{n},\bm{m}}({\xi}_j \hat{e}_j) &= \mathcal{P}_{\bm{n},\bm{m}}(\bm{0}) +2\xi_j \left( \sqrt{n_j}\,\text{Re}[ {C}_{\bm{n},\bm{m}} {C}^\ast_{{\bm{n}-\hat{e}_j},\bm{m}} ] - \sqrt{n_j+1} \,\text{Re}[ {C}_{\bm{n},\bm{m}} {C}^\ast_{{\bm{n}+\hat{e}_j},\bm{m}}] \right) \nonumber\\
    \mathcal{P}_{\bm{n},\bm{m}}(i{\xi}_j\hat{e}_j)&=\mathcal{P}_{\bm{n},\bm{m}}(\bm{0})+ 2\xi \left( -\sqrt{n_j}\,\text{Im}[ {C}_{\bm{n},\bm{m}} {C}^\ast_{{\bm{n}-\hat{e}_j},\bm{m}}]+\sqrt{n_j+1}\,\text{Im}[ {C}_{\bm{n},\bm{m}} {C}^\ast_{{\bm{n}+\hat{e}_j},\bm{m}}] \right) \ ,
\end{align}
where $\hat{e}_j$ is the unit vector along component $j$.

For a $2L$-mode system with maximum PNR resolution of $K$ photons on each mode, displacing in this way leads to $4L+1$ measurement settings which give rise to $(4L+1)^K$ equations that can be solved to determine the $(4L)^K$ real parameters associated with the possible correlation functions. However, not all correlators are needed.

Next, we discuss how to 
extract the relative phase of correlation functions at different times, using the above results. By differentiating with respect to time and inserting a complete set of Fock states, we obtain
\be \frac{d C_{\bm{n}, \bm{m}} (t)}{dt} = i \sum_{\bm{n}'} \brasub{\bm{n}}{\gamma} H \ketsub{\bm{n}'}{\gamma} C_{\bm{n}', \bm{m}} (t) \ee
via which one can access the time-dependent global phase that could not be determined through Eq.\ \eqref{eq:75}.

Next, we discuss two alternative methods of obtaining the global phase with quantum computation.
Suppose we wish to relate the global phases at times $t_a, t_b$ with $t_b > t_a > 0$.

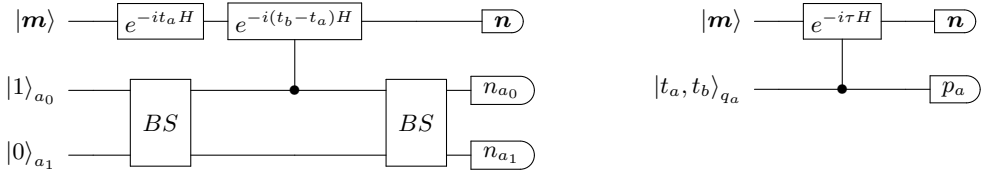
\begin{figure}
    \centering
\[	\Qcircuit @C=1em @R=1.5em {
		\lstick{\ket{\bm{m}}} &\qw&\gate{e^{-it_aH}}			   &\gate{e^{-i(t_b-t_a)H}} &\qw&\measureD{\bm{n}}\\
		\lstick{\ket{1}_{a_0}}   &\qw&\multigate{1}{BS}&\ctrl{-1}		 &\multigate{1}{BS}&\measureD{n_{a_0}}\\
		\lstick{\ket{0}_{a_1}}   &\qw&\ghost{BS}       &\qw              &\ghost{BS}&\measureD{n_{a_1}}				 
	} \quad\quad\quad\quad\quad\quad\quad\quad\quad  \Qcircuit @C=1em @R=1.5em {
		\lstick{\ket{\bm{m}}} &\qw			   &\gate{e^{-i\tau H}}&\qw&\measureD{\bm{n}} \\
		\lstick{\ket{t_a,t_b}_{q_a}}   &\qw&\ctrl{-1}		 &\qw&\measureD{p_a}
	} \]
    \caption{Quantum circuits for the calculation of the time-dependent global phase of the one-point correlation function \eqref{eq:60} using photon-number and quadrature control.}
    \label{fig:Cirq_TwoTimes}
\end{figure}

The first method relies on photon-controlled evolution.
We prepare our state $\ket{\bm{m}}$ and evolve it with $e^{-it_a H}$. We attach two ancillary qumodes, labeled as $a_0$ and $a_1$, one in a single photon state, $\ket{1}_{a_0}$, and the other in the vacuum state, $\ket{0}_{a_1}$. These are interfered in a 50/50 beam splitter to generate the entangled state $(\ketsub{01}{a_0a_1}+\ketsub{10}{a_0a_1})/\sqrt2$. Mode $a_0$ is used as control for the evolution of our state by $e^{-i(t_b-t_a) H}$. This requires the photon-number-controlled gate
\be\label{eq:ControlledNHamiltonian} CU_{a_0} \equiv \exp\{ -i (t_b- t_a) N_{a_0} \otimes H \} \ ,\ee 
where $N_{a_0}$ is the photon number of the ancilla qumode $a_0$. This is a non-Gaussian gate that increases the complexity of the quantum simulation. The ancilla modes interfere again at a second beam splitter and their photon numbers are measured. Only one detector is needed, because there is only one photon between the two ancilla modes. This is equivalent to a Mach-Zehnder interferometer for the ancilla modes where one of the arms is used as a control for the gate \eqref{eq:ControlledNHamiltonian}. The quantum circuit for a single mode (Eq.\ \eqref{eq:60}), for simplicity in notation, is shown in Fig.\ \ref{fig:Cirq_TwoTimes}.
Explicitly, the state in the quantum circuit before the measurements  is given by
\begin{align}
	\ket{\text{out}} &= BS_{a_0a_1}\cdot e^{-it_a H} \cdot CU_{a_0} \cdot BS_{a_0a_1} \ket{\bm{m}}\ketsub{1}{a_0}\ketsub{0}{a_1} \nonumber\\
	&= \frac{1}{2}\left[ \left( e^{-i t_bH}  + e^{-it_a H} \right) \ket{\bm{m}}  \ketsub{1}{a_0}\ketsub{0}{a_1}+ \left( e^{-i t_bH}  - e^{-it_a H} \right) \ket{\bm{m}} \ketsub{0}{a_0}\ketsub{1}{a_1}\right]
\end{align}
The only possible photon number measurement outcomes on the ancilla qumodes are $(n_{a_0},n_{a_1}) \in \{ (1,0) ,( 0,1) \}$. After measuring all modes, we obtain the outcome $(\bm{n}, n_{a_0}, n_{a_1})$ with probability
\begin{align}\label{eq:PhaseDiff1}
\mathcal{P}_{\bm{n},n_{a_0}}  &= \frac{1}{4} \left| C_{\bm{n}, \bm{m}} (t_a) + (-)^{n_{a_0}} C_{\bm{n}, \bm{m}} (t_b) \right|^2 \nonumber\\
&= \frac{1}{4} \left[ |C_{\bm{n}, \bm{m}} (t_a)|^2+ |C_{\bm{n}, \bm{m}} (t_b)|^2 +2(-)^{n_{a_0}} \cos\left( \Phi(t_b)-\Phi(t_a) \right) |C_{\bm{n}, \bm{m}} (t_a)C_{\bm{n}, \bm{m}} (t_b)| \right]
\end{align}
showing explicitly the dependence on the global phase $\Phi (t)$.
Having already obtained $|C_{\bm{n}, \bm{m}} (t)|$, we can calculate the relative phase $\Phi(t_b)-\Phi(t_a)$  with experimental data using Eq.\ \eqref{eq:PhaseDiff1}.

An alternative method is based on field-controlled evolution. In this case, we attach a single ancilla qumode labeled $a$ in a state that consists of two sharp peaks at the quadrature values $q_a = s_a, s_b$, denoted $\ketsub{s_a,s_b}{q_a}$. This assumes that we have access to such a state, such as a squeezed Schr\"odinger cat state, which is challenging to engineer. We use the ancilla quadrature as control in the evolution of the state $\ket{\bm{m}}$, by applying the non-Gaussian quadrature-controlled gate
\be CU_q \equiv \exp \{ -i \tau q \otimes H \} \ee
which is less complex than the photon-number-controlled gate \eqref{eq:ControlledNHamiltonian}. The parameters are chosen so that $t_a = s_a\tau$, $t_b = s_b \tau$. Finally, we perform a homodyne detection measuring the $p$ quadrature of the ancilla. The quantum circuit is shown in Fig.\ \ref{fig:Cirq_TwoTimes}.

Measurement of the output state gives us access to the probability distribution
\begin{eqnarray}
\mathcal{P} (p) &\propto&	\left| {C}_{\bm{m}, \bm{n}} (t_a) e^{ips_a}+{C}_{\bm{m}, \bm{n}} (t_b) e^{ips_b} \right|^2 \nonumber\\
&\approx& |C_{\bm{n}, \bm{m}} (t_a)|^2+ |C_{\bm{n}, \bm{m}} (t_b)|^2 + \cos\left( \Phi(t_b)-\Phi(t_a) + p(s_b - s_a) \right) |C_{\bm{n}, \bm{m}} (t_a)C_{\bm{n}, \bm{m}} (t_b)| 
\end{eqnarray}
from which the phase difference $\Phi(t_b)-\Phi(t_a)$ can be calculated.

\subsection{Three-point correlation functions}

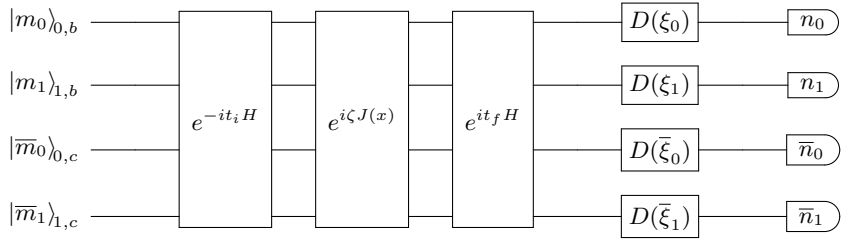
\begin{figure}[t]
    \centering
    \[\Qcircuit @C=1.8em @R=1em { \lstick{\ketsub{m_0}{0,b}} & \qw  & \multigate{3}{e^{-it_iH}}  & \multigate{3}{e^{i\zeta J(x)}}  & \multigate{3}{e^{it_fH}}  &\qw  & \gate{D(\xi_0)} &   \qw   & \measureD{n_0} \\  \lstick{\ketsub{m_1}{1,b}} & \qw  & \ghost{e^{-it_iH}} & \ghost{e^{i\zeta J(x)}}  & \ghost{e^{it_fH}}& \qw  & \gate{D(\xi_1)} &  \qw   & \measureD{n_1} \\ \lstick{\ketsub{\overline{m}_0}{0,c}} & \qw  & \ghost{e^{-it_iH}} & \ghost{e^{i\zeta J(x)}}  & \ghost{e^{it_fH}} & \qw  & \gate{D(\overline{\xi}_0)} & \qw   & \measureD{\overline{n}_0} \\ \lstick{\ketsub{\overline{m}_1}{1,c}}  & \qw  & \ghost{e^{-it_iH}} & \ghost{e^{i\zeta J(x)}}  & \ghost{e^{it_fH}} & \qw  & \gate{D(\overline{\xi}_1)}   & \qw    & \measureD{\overline{n}_1} } \]
    \caption{Phase-sensitive simulation of the three-point correlator \eqref{eq:85} for two lattice sites requiring four qumodes.}
    \label{fig:disp3}
\end{figure}

Beyond simulating Hamiltonian evolution, we seek matrix elements of currents, such as given by \eq{C3pt_def}. Consider the three-point correlation function
\be\label{eq:85} {C}_{\bm{n},\bm{m}} (t_i, t_f) = \bra{\bm{n}}  e^{it_f H} \mathcal{J} (x) e^{-it_i H} \ket{\bm{m}} \ee
where $\mathcal{J} (x)$ is the $U(1)$ current,
\be\label{eq:J} \mathcal{J} (x) = i \left[ \phi(x) \pi^\dagger(x) - \phi^\dagger(x) \pi(x) \right] \ee
To compute the three-point correlator, we work as before by inserting displacement operators and taking derivatives. Additionally, we insert
the Gaussian unitary $e^{i\zeta \mathcal{J} (x)}$,
instead of $\mathcal{J} (x)$ itself, which is a Hermitian operator, and expand for small $\zeta$, $e^{i\zeta \mathcal{J} (x)} = \mathbb{I} + i\zeta J(x) + \mathcal{O} (\zeta^2)$, to recover the desired amplitude. The quantum circuit for the calculation of the three-point correlator on two lattice sites requiring four qumodes is shown in Fig.\ \ref{fig:disp3}.
This procedure yields the three-point correlation functions up to a global phase. To calculate the phase, we work as before to express the derivative with respect to time $t_f$ as
\be \frac{d C_{\bm{n}, \bm{m}} (t_i,t_f)}{dt_f} = i \sum_{\bm{n}'} \brasub{\bm{n}}{\gamma} H \ketsub{\bm{n}'}{\gamma} C_{\bm{n}', \bm{m}} (t_i, t_f) \ee
from which we can deduce the global phase as a function of time.

Alternatively, we can introduce ancilla qumodes to perform a controlled evolution by controlling either with photon number or a quadrature, similar to the process depicted in Fig.\ \ref{fig:Cirq_TwoTimes}.

It is straightforward to extend this method to higher-point correlators of currents, albeit it becomes more cumbersome as the number of insertions of currents increases.

\section{Measurement-based quantum optical simulation}
\label{secn:mbqc}
In this section, we turn to MBQC implementation, which allows us to leverage the record scalability achieved in the generation of cluster states~\cite{Yokoyama2013,Chen2014,Yoshikawa2016,Asavanant2019,Larsen2019}, the quantum substrates for universal quantum computing~\cite{Briegel2001,Raussendorf2001}.

In addition, an MBQC implementation allows us to achieve two critical simulation steps: \textit{(i)}, the generation of arbitrary input Fock states (which map directly to excitation states of the free-field QFT) and \textit{(ii)}, the \textit{deterministic} generation of the nonunitary polynomial gates which simulate current matrix elements. Nonunitary quantum gates in particular are very nonstandard and require non-Gaussian quantum optical resources. It has been shown that the formalism of measurement-based quantum computing and, in particular, its photonic implementation using continuous-variable cluster states, can be leveraged to great effect by the simple addition of photon-number-resolved (PNR) measurements~\cite{Eaton2022_PhANTM}.

\subsection{Input Fock States}
\label{secn:f}

As presented in \secn{2}, eigenstates of the free Hamiltonian $H_0$ are products of states in the photon-number basis, $\ket{n}_\gamma$. The states can be created within the cluster by performing PNR detection on cluster state nodes (See Appendix \ref{AppA}). From a quantum optical viewpoint, this can be seen immediately if one thinks of a two-mode cluster state as a locally rotated two-mode squeezed state $\sum_{n=0}^\infty(\tanh r)^n/\cosh r \,\ketsub n1\ketsub n2$, which is perfectly photon-number correlated. Hence PNR detecting $n_o$ photons in one mode will result in projecting the other mode onto Fock state $\ket {n_o}$. This can also be shown in the Gaussian teleportation gate formalism whose homodyne detector is replaced by a PNR detector in order to project on the Fock basis. First, we write a Fock state in the quadrature eigenbasis,
\begin{align}
    \ket{n}_\gamma&=(2\pi)^{-\frac12}\int^\infty_{-\infty}\!ds\,\psi_n(s)\ketsub{s}{\gamma_q}, 
\end{align}
where $\psi_n(s)$ is the well-known Hermite-Gauss wavefunction, normalized here to $\sqrt{2\pi}$ for mere convenience
\begin{align}
   \psi_n(s)&=(2\pi)^{\frac12}\times\pi^{-\frac14}(n!\,2^n)^{-\frac12}\,H_n(s)\,e^{-\frac{s^2}2},
\end{align}
$H_n(s)$ being the Hermite polynomial of order $n$. We then re-express the Fock state as a quadrature eigenstate modified by the non-Gaussian operator $\psi_n(Q)$:
\begin{align}
    \ket{n}_\gamma    &=(2\pi)^{-\frac12}\,\psi_n(Q)\int^\infty_{-\infty}\!ds\,\ketsub{s}{\gamma_q} \\
\ket{n}_\gamma     &= \psi_n(Q)\,\ketsub{0}{\gamma_p}.
\end{align}
Now, we replace the homodyne detector in the canonical teleportation circuit of \eq{canon_telep} by a PNR detector as operator $F_n$ and use the previous result\footnote{note that, unlike previous figures, these circuits flow from right to left to better reflect operatorial order.}
\begin{equation}
\raisebox{-1.25em}{$F_n$\ =\hspace{15mm}}\Qcircuit @C=1em @R=2em {
\lstick{\brasub{n}{\gamma1}} &\qw& \ctrl{1} & \rstick{(\text{in})_1} \qw \\
\lstick{(\text{out})_2} &\qw&\control \qw &\rstick{\ketsub{0}{\gamma_p2}} \qw
}\raisebox{-1.25em}{\hspace{15mm}$=$\hspace{15mm}}
\Qcircuit @C=1em @R=2em {
\lstick{\brasub{0}{\gamma_p1}} &\gate{\psi_n(Q_1)}& \ctrl{1}  &\qw&\rstick{\text{(in)}_1} \qw \\
\lstick{\text{(out)}_2} &\qw&\control \qw &\qw&\rstick{\ketsub{0}{\gamma_p2}}\qw
}\raisebox{-1.25em}{\hspace{15mm},}
\end{equation}
where the application of non-Gaussian gate $\psi_n(Q_1)$ allows us to formally restore quadrature detection on mode 1. Since the CZ gate $\exp(iQ_1Q_2)$ contains only amplitude quadrature operators it commutes with $\psi_n(Q_1)$, yielding
\begin{equation} 
\raisebox{-1.25em}{$F_n$\ =\hspace{15mm}}\Qcircuit @C=1em @R=2em {
\lstick{\brasub{n}{\gamma1}} &\qw& \ctrl{1} & \rstick{(\text{in})_1} \qw \\
\lstick{(\text{out})_2} &\qw&\control \qw &\rstick{\ketsub{0}{\gamma_p2}} \qw
}\raisebox{-1.25em}{\hspace{15mm}$=$}
\hspace{15mm}
    \Qcircuit @C=1em @R=2em {
\lstick{\brasub{0}{\gamma_p1}}&\qw & \ctrl{1} &\gate{\psi_n(Q_1)}&\rstick{\text{(in)}_1} \qw \\
\lstick{\text{(out)}_2} &\qw&\control \qw &\qw&\rstick{\ketsub{0}{\gamma_p2}} \qw
}\raisebox{-1.25em}{\hspace{15mm},}
\end{equation}
which is the familiar teleportation gate of \eq{canon_telep} and therefore results in the measurement-based operator (with the now familiar $2\leftarrow1$ swap implied) 
\begin{equation}\label{eq:pnr_telep}
    F_n=R(\tfrac{\pi}{2})\,\psi_n(Q).
\end{equation}
If input $\text{(in)}_1$ is a bare cluster state node $\ketsub{0}{1P}$, then $F_n\ketsub{0}{1P}=i^n\ket{n}_{2\gamma}$, i.e., a Fock state to a global phase left. Thus, performing PNR detection on the cluster state allows one to embed Fock states within the cluster. 

Crucially, one can see that PNR directly enacts a nontrivial ``Hermite-Gaussian'' non-Gaussian gate which contains an $n^\text{th}$-order polynomial in $Q$. We will see in \secn{Iexp} that a variant of PNR detection, called photon subtraction, on the CV cluster state allows us to separate out the polynomial gate in $Q$~\cite{Eaton2022_PhANTM}, which  is required for the evaluation of current matrix elements discussed in \secn I, thereby yielding a unique capability of our PNR-augmented CV cluster state paradigm. 

Finally, note that, although the PNR detection results, i.e., the input photon numbers, are stochastic, one can always apply PNR detection to several cluster state nodes simultaneously and subsequently re-route the desired outcome once it is obtained, by teleportation to the desired cluster state location through homodyne measurements. This parallel ``repeat-until-success'' method does not carry a significant overhead and allows precise state preparation. 

So far, we have described how CV cluster state and PNR detection can be used together to apply an arbitrary Gaussian unitary $G$ to a well-defined Fock-state input, amounting, respectively, to well-defined particle numbers in free-field and momentum eigenmodes. Regarding quartic-phase gates, exact expressions in terms of (15) cubic gates and Gaussian gates have been derived~\cite{Kalajdzievski2019}, as was mentioned earlier. These expressions can be straightforwardly translated into MBQC operations over cluster states~\cite{Gu2009} seeded by cubic-phase states. We now turn to the implementation of the current matrix elements.

\subsection{Optical quantum simulation of current operators}\label{secn:Iexp}

Beyond simulating Hamiltonian evolution, we seek matrix elements of currents, such as given by \eq{C3pt_def}. In this case, generic currents given by $\mathcal{J}$ take the form of polynomials in the field operators. We know show that such operator insertions in correlation functions can be achieved through the ability to perform the non-unitary operation of photon subtraction on the cluster state. This, when iterated, allows one to apply polynomial gates of arbitrary degree via gate teleportation~\cite{Eaton2022_PhANTM}.

\subsubsection{Polynomial gates}

To demonstrate how this can be achieved, we can modify the teleportation circuit shown in \eq{canon_telep} by applying the photon subtraction operator immediately prior to the homodyne measurement and fixing the measurement angle to $\theta=0$. Photon subtraction can be experimentally achieved by placing a weakly reflective beamsplitter, of reflection coefficient $r\to0$,  
in the path of the light and sending the reflected portion to a PNR detector. For a PNR detection outcome of $n$ photons, the subtraction operator is given by 
\begin{equation}
\mathcal{S}_n= 
(-1)^n(n!e^{n\beta})^{-\frac12}
\left(2\sinh{\beta}\right)^{n/2}e^{-\beta a^\dag a}\,a^n,
    \label{eq:kraus_photsubt}
\end{equation}
where $\beta=-\frac12\ln{1-r^2}$.  
For a vanishing beamsplitter reflectivity $r\to0$, $\beta\to0$ and the photon subtraction becomes a nearly ideal application of the photon-downshift operator: $\mathcal{S}_n\rightarrow (n!)^{-\frac12}\,a^n$. Considering this limiting case leads to a circuit representation given by
\begin{equation}\label{eq:canon_subtr}
    \Qcircuit @C=1em @R=1.5em {
\lstick{\brasub{m}{\gamma_p}} &\qw&\gate{a^n} &\ctrl{1} & \rstick{(\text{in})} \qw \\
\lstick{(\text{out})} &\qw&\qw&\control \qw &\rstick{\ketsub{0}p} \qw
}\raisebox{-1em}{\hspace{10mm},}
\end{equation}
which has a Kraus operator representation given by~\cite{Eaton2022_PhANTM}
\begin{equation}
    K_n\propto X(m)R(\tfrac{\pi}{2})H_n\left(\frac{iQ-m}{\sqrt{2}}\right),
    \label{eq:limit_case_final}
\end{equation}
where $H_n(x)$ are the Hermite polynomials of degree $n$. It should be noted that using a small photon-subtraction beamsplitter reflectivity comes at the cost of decreasing the success probability of a subtraction event; however, failing to subtract any photons will reduce the applied operator to $K_0\propto X(m)R(\tfrac{\pi}{2})$, which simply acts as the regular teleportation measurement. Although it was shown in that the repeated application of operators of the form of $K_n$ can lead to generating and embedding Schr\"odinger cat states in a cluster state, even under imperfect conditions including Gaussian noise from finite squeezing and away from the vanishing beamsplitter reflectivity limit~\cite{Eaton2022_PhANTM}, we can also use this process to implement polynomial gates of arbitrary degrees.

By taking advantage of the ability to perform changes to the basis of a future measurement with feedforward operations, and by performing a regular teleportation following the circuit in \eq{canon_subtr}, we can enact the operator $K'_n\propto H_n\left(\frac{iQ-m}{\sqrt{2}}\right)$. As we cannot control the detection result, the specific Hermite polynomial in $Q$ that is applied will be stochastic, but \textit{known}, as $m$ is a homodyne measurement result and $n$ is a PNR measurement result. 

Note that we can also tune the reflectivity of the beamsplitter used in photon subtraction to be small enough such that cases where $n\geq 2$ photons are detected is vanishingly small. In this case, we will only ever apply the operator $K'_1\propto Q+im$ or $K'_0 \propto \mymathbb{1}$. Thus, by repeating this circuit until the operator $K'_1$ has been applied $k$ times, we can controllably implement an operator that is a degree $k$ polynomial in the quadrature operator, $Q$. This final operator could be used to represent a current, and would be
\begin{equation}
    \mathcal{J}=\Pi^k_{j=1}(Q+im_j),
    \label{eq:poly_gate}
\end{equation}
where $m_j\in \mathbb{R}$ is the homodyne measurement result of the $j$-th teleportation circuit where photon subtraction was successful.

\subsubsection{Gate Teleportation}
The polynomial expressed in \eq{poly_gate} may not always be the desired operation as one cannot control the measurement results, $m_j$ at each step. However, if one has a sufficiently large cluster state, regions can be used to generate a polynomial resource ahead of time and use it to apply a desired gate later through gate teleportation. Suppose that one began with several bare cluster state nodes, each a zero-momentum eigenstate, $\ketsub{0}{\gamma_p}$, and iterated the method outline above on each node separately until the measurement results $m_j$ were near enough to a desired polynomial such that one of the states was transformed to 
\begin{equation}
\ket{\psi'}=f(Q)\ketsub{0}{\gamma_p},    
\end{equation}
where $f(Q)$ is the polynomial gate we wish to apply. Now, by teleporting the resource $\ket{\psi'}$ across the cluster state until it is neighboring the quantum information that we wish to apply the gate to, $\ket{\psi}$, and first applying a rotation $R^\dag(\tfrac{\pi}{2})$ to $\ket{\psi}$, we can implement the circuit
\begin{equation}\label{eq:gen_telep}
\begin{split}
\raisebox{-1.5em}{$\ketsub{\psi}{out}=$}
\hspace{10mm}
\Qcircuit @C=1em @R=2em {
\lstick{\brasub{m}{\gamma_p}} &\qw& \ctrl{1}  &\gate{R^\dag(\tfrac{\pi}{2})}&\rstick{\ket{\psi}} \qw \\
\lstick{} &\qw&\control \qw &\qw&\rstick{\ket{\psi'}}\qw
}\raisebox{-1.5em}{\hspace{10mm}$=$}
\hspace{15mm}
    \Qcircuit @C=1em @R=1.25em {
\lstick{\brasub{0}{\gamma_p}}&\qw &\qw& \ctrl{1} &\gate{Z^\dag(m)}&\gate{R^\dag(\tfrac{\pi}{2})} &\rstick{\ket{\psi}} \qw \\
\lstick{} &\qw&\gate{f(Q)}&\control \qw &\qw&\qw&\rstick{\ketsub{0}{\gamma_p}} \qw
}\raisebox{-1.5em}{\hspace{8mm}.}
\end{split}
\end{equation}
Pulling the circuit taut and commuting the resultant rotation operation to the end, we get
\begin{align}
    \ketsub{\psi}{out}&=f(Q)R(\tfrac{\pi}{2})Z^\dag(m)R^\dag(\tfrac{\pi}{2})\ket{\psi} \\
    &=f(Q)X(m)\ket{\psi}.
\end{align}
Thus, we can use gate teleportation to apply any operator in $Q$, including polynomial gates, up to a Gaussian measurement-dependent shift on the input quantum information.

\section{Conclusion} 
\label{secn:conc}

In this paper, we have proposed a detailed program, both theoretical and experimental, for computing observables in quantum field theories non-perturbatively using a single analog photonic quantum processor. In particular, in Sec.~\ref{secn:obs}, we reviewed a class of time-dependent nuclear physics observables that are currently inaccessible via lattice QCD implemented on classical computers.

Our proposal has several unique, crucial features:
\begin{itemize}
    \item[\textit{(i)}] it allows phase-sensitive reconstruction of scattering amplitudes, beyond merely reconstructing a probability distribution by sampling from it;  \item[\textit{(ii)}] it leverages the massive scalability of CV cluster-state implementation, in particular in the quantum optical frequency comb~\cite{Pysher2011,Chen2014,Pfister2019} but also in the time domain~\cite{Yokoyama2013,Yoshikawa2016,Asavanant2019,Larsen2019} and also of hybrids of the two~\cite{Alexander2016a}, to envision scalable implementations of QFT simulation on a single photonic quantum processor; \item[\textit{(iii)}] finally, it leverages the newly discovered capability of CV cluster states to efficiently implement non-Gaussian, as well as non-unitary, quantum gates such as current terms~\cite{Eaton2022_PhANTM}. 
\end{itemize}
All of this bodes well for, ultimately, the experimental feasibility of the quantum simulation of quantum field theory beyond current lattice-gauge QCD capabilities. Although we are not at a stage yet to consider the complexities of QCD using the proposed setup, the path towards being able to study QCD with a photonic quantum processor is becoming clearer.

\acknowledgments

We thank Ignacio Cirac for useful comments on quantum simulation overheads. This work was supported by the Jefferson Lab LDRD project No. LDRD21-17 under which Jefferson Science Associates, LLC, manages and operates Jefferson Lab. RGE acknowledges support from the U.S. Department of Energy contract DE-AC05-06OR23177, under which Jefferson Science Associates, LLC, manages and operates Jefferson Lab, and the Department of Energy, Office of Science, National Information Science Research Centers, Co-design Center for Quantum Advantage (C$^2$QA) under contract number DE-SC0012704. OP acknowledges support from National Science Foundation grants PHY-2112867 and ECCS-2219760. GS acknowledges support by the National Science Foundation under award DGE-2152168,  the Army Research Office under award W911NF-19-1-0397, and the Department of Energy under awards DE-SC0023687 and DE-SC0024325.


\appendix

\section{Review of photonic one-way quantum computing}
\label{AppA}
In this appendix we recall fundamentals of measurement-based universal quantum computing with cluster states~\cite{Briegel2001}, a.k.a.\ one-way quantum computing~\cite{Raussendorf2001}. We focus on the continuous-variable photonic implementation~\cite{Menicucci2006,Menicucci2008,Menicucci2014ft} in which quantum optical fields, a.k.a.\ qumodes, are employed in lieu of qubits.
As quantum computing has been formulated for CV encodings~\cite{Lloyd1999,Gottesman2001,Bartlett2002}, so has MBQC~\cite{Zhang2006,Menicucci2006,vanLoock2007,Gu2009}. Center to this endeavor has been the experimental realization of record size qumode-based cluster states with the most scalability promise across any quantum computing platform, be it qubit- or qumode-based~\cite{Pfister2004,Menicucci2008,Pysher2011,Chen2014,Yokoyama2013,Yoshikawa2016,Asavanant2019,Larsen2019}. See Ref.~\cite{Pfister2019} for a review.

\subsection{Basics of quantum optics}

\subsubsection{Quantum optical field}
The quantized electromagnetic plane-wave field has the expression, in the Heisenberg picture,
\begin{align}\label{eq:E}
E(\vec r,t)  \propto a e^{i(\vec k\cdot\vec r-\omega t)}+a^\dag e^{-i(\vec k\cdot\vec r-\omega t)}
= Q \cos(\vec k\cdot\vec r-\omega t) -P \sin(\vec k\cdot\vec r-\omega t),
\end{align}
where $a$ and $a^\dag$ are the photon annihilation and creation operators, $Q=(a+a^\dag)/\sqrt2$ is the position-like amplitude quadrature, and $P=i(a^\dag-a)/\sqrt2$ is the momentum-like phase quadrature.

Continuous-variable quantum information uses the amplitude and phase quadrature eigenstates, $\ketsub s{\gamma_q},\ketsub s{\gamma_p}$, which satisfy
\begin{align}
    Q\ketsub s {\gamma_q} = s\ketsub s{\gamma_q}\nonumber\\
    P\ketsub s {\gamma_p} = s\ketsub s{\gamma_p},
\end{align}
and are Fourier transforms of each other. Because of the Heisenberg inequality $\Delta Q\Delta P\geq 1/2$, these states have infinite energy and are therefore unphysical. 

\subsubsection{Single-mode squeezing} 
In the laboratory, arbitrarily good approximations of the quadrature eigenstates are realized by squeezed states, obtained from two-photon emitters. Indistinguishable pair photons yield single-mode squeezed states of the form \begin{align}\label{eq:sms1}
S(r)\ketsub 0\gamma&=e^{\frac r2(a^{\dag2}-a^2)}{\ketsub 0\gamma}\\
&\propto \sum_{n=0}^\infty \psi_n\,\ketsub{2n}\gamma\propto \int ds\,e^{-\frac{s^{2}}{2e^{2r}}} \,\ketsub s{\gamma_q}\propto \int ds\,e^{-\frac{s^{2}}{2e^{-2r}}} \,\ketsub s{\gamma_p}\label{eq:sms2}
\end{align}
where $S(r)$ is the squeezing operator and $r$ the squeezing parameter, related to the two-photon emission rate~\cite{Walls1994}. For $r>0$ the state is phase squeezed, else it's amplitude squeezed. The limit $r\to\pm\infty$, corresponding to infinite energy, yields unphysical ``plane wave'' phase-quadrature eigenstates. 

\subsubsection{Two-mode squeezing} 
Distinguishable pair photons yield two-mode squeezed states, which are entangled states of the form 
\begin{align}
S(1,2;r)\ketsub 0\gamma&=e^{r(a_1^{\dag}a_2^{\dag}-a_1a_2)}\ketsub 0\gamma\label{eq:tms}\\
&\propto\sum_{n=0}^\infty \phi_n\,\ketsub n{\gamma1}\ketsub n{\gamma2}\propto\iint ds ds'\,e^{-\frac{(s-s')^2}{2e^{-2r}}}\,e^{-\frac{(s+s')^2}{2e^{2r}}}\ketsub s{\gamma_q1}\ketsub {s'}{\gamma_q2}\propto\iint ds ds'\,e^{-\frac{(s+s')^2}{2e^{-2r}}}\,e^{-\frac{(s-s')^2}{2e^{2r}}}\ketsub s{\gamma_p1}\ketsub {s'}{\gamma_p2}.\label{eq:tms2}
\end{align}
and are the closest physical approximations of the unphysical Einstein-Podolsky-Rosen (EPR) entangled states~\cite{Einstein1935}
\begin{align}
\ketsub {\text{EPR}}{\gamma12}=\sum_{n=0}^\infty \ketsub n{\gamma 1}\ketsub n{\gamma 2}=\int ds \,\ketsub s{\gamma_q1}\ketsub {s}{\gamma_q2}=\int ds \,\ketsub s{\gamma_p1}\ketsub {-s}{\gamma_p2}.\label{eq:epr}
\end{align}
In the Heisenberg picture, the two-mode squeezing process is described by the following Bogoliubov transformation
\begin{align}
a_1(r) &= S(1,2;r)^\dag\, a_1 \,S(1,2;r) = a_1\,\cosh r + a_2^\dag\,\sinh r \nonumber\\
a_2(r) &= S(1,2;r)^\dag\, a_2\, S(1,2;r) = a_2\,\cosh r + a_1^\dag\,\sinh r.\label{eq:bogo1}
\end{align}

\subsection{CV cluster states}

Over qubits, cluster states (a.k.a.\ graph states) are constituted of controlled-Pauli-Z gates linking qubits all initialized in the $\ket+=(\ket0+\ket1)/\sqrt2$ state. Over qumodes (quantum fields), the $\ket+$ states correspond, ideally, to $p=0$ quadrature eigenstates. On a lattice of $L$ sites, we need $L$ qumodes for particles (labeled $(k,b)$) and an equal number for anti-particles (labeled $(k,c)$) of physical momentum $\frac{2\pi k}{L}$ ($k\in \mathbb{Z}_L$). CV cluster states are then formed by entangling qumodes in $p=0$ phase-quadrature eigenstates $\ketsub{0}{\gamma_pk,b} = (2\pi)^{-1/2}\int ds \, \ketsub {s} {\gamma_qk,b}$, (and similarly for an anti-particle created by $c^\dag (k)$) or, in the laboratory, phase-squeezed states $S(k,b;r) \ketsub{0}{k,b}= \exp\{\frac{r}{2} [ b^{\dagger 2} (k) - b^2 (k)]\}\ketsub{0}{\gamma k,b}$, with controlled-$Z$ entangling gates ${CZ}=e^{iQ_k \otimes Q_l}$. For the sake of simplicity, we retain quadrature eigenstates for now. A cluster state graph is defined with the aforementioned qumodes as vertices linked by CZ edges. A two-mode example is
\begin{equation}\label{eq:C}
\setlength{\unitlength}{.35in}
\begin{picture}(8,1.1)
\thicklines
\put(-1.1,.5){$e^{iQ_k  Q_l}\,\ketsub{0}{\gamma_p,k,b} \ketsub{0}{\gamma_p,l,b} \ \  =$}
\put(4,0.6){\circle*{.4}}
\put(6,0.6){\circle*{.4}}
\put(4,0.6){\line(1,0){2}}
\end{picture}
\end{equation}
This state is identical to the EPR state, to a local (single-qumode) Fourier transform left.

In the MBQC paradigm, the universal QC gate set can be implemented by single qubit measurements and feedforward to nearest neighbors on a 2D cluster graph. In CV MBQC, these measurements fall into two categories: field, i.e., quadrature measurements and PNR measurements. Field measurements project onto the aforementioned quadrature eigenstates, $\{ \ketsub{s}{\gamma_q}\}_{s\in\mathbb R}$, $\{ \ketsub{s}{\gamma_p}\}_{s\in\mathbb R}$, . These measurements are implemented by homodyne detection, i.e., interference with a well-calibrated and stable local oscillator (LO) laser whose phase relative to that of the measured quantum fields determines which of its quadratures is measured. PNR measurements project onto the discrete Fock basis states $\{ \ketsub n\gamma\}_{n\in\mathbb N}$.

In terms of MBQC gates, these two types of measurements have fundamentally distinct capabilities: on one hand, quadrature-measurement-enabled MBQC gates are Gaussian in nature (in terms of their Wigner function) and therefore equivalent to Clifford operations on qubits~\cite{Gottesman2001,Bartlett2002}. We present this in more detail later. On the other hand, PNR measurements give access to non-Gaussian gates, whose qubit counterparts are non-Clifford gates, and enable exponential speedup~\cite{Gottesman1999a,Bartlett2002}, as well as quantum error correction~\cite{Gottesman2001}. In this work, PNR measurements are key to initial Fock-state preparation (thereby defining the initial excitation number of free fields), non-Gaussian  ($e^{-it H_{\text{int}}}$) gate implementation, and final measurements of free field excitations after  evolution involving quartic phase gates. 

It is important to note that PNR measurements of CV cluster states have recently been discovered as a powerful enabling paradigm for MBQC~\cite{Eaton2022_PhANTM} and have the potential to do the same for quantum simulation.

\subsection{Basic principles of CV MBQC}

We start by recalling the fundamentals of CV MBQC as they relate to cluster entangled states. Details can be found in Refs.~\cite{Gu2009} and \cite{Pfister2019}. 
A central QC primitive is the teleportation gadget, which can be expressed in circuit form as  
\begin{equation}
\label{eq:canon_telep}
T_{2\leftarrow1}\ =\ \brasub{m}{\gamma_\theta1}{CZ}_{12}\ketsub{0}{\gamma_p2}\quad = \qquad\qquad 
    \Qcircuit @C=1.5em @R=1.2em {
\lstick{\brasub{m}{\gamma_\theta1}} &\qw& \ctrl{1} & \rstick{(\text{in})_1} \qw \\
\lstick{(\text{out})_2} &\qw&\control \qw &\rstick{\ketsub{0}{\gamma_p2}} \qw
}\raisebox{-.5em}{\hspace{8mm}\quad,}
\end{equation}
where the vertical line linking the two dots denotes the $CZ_{12}$ gate---note that, if $(\text{in})_1=\ketsub0{\gamma_p1}$, then the input is the canonical cluster state of \eq C, and where the top left bra denotes a measurement, with result $m$, of the $\theta$-rotated quadrature $P_\theta=R^\dag(\theta)PR(\theta)=-Q\sin\theta+P\cos\theta$, where $R(\theta)=\exp(-i\theta a^\dag a)$. Note that this circuit proceeds from right to left so that one can straightforwardly write down the circuit operators in the order that they apply. Any quantum state, potentially multi-mode (which will obviously increase the circuit breadth), can enter the circuit at placeholder $(\text{in})_1$ and come out teleported---and possibly acted on by a gate---at placeholder $(\text{out})_2$. For teleportation, $(\text{in})_1=\ket\psi_1$ and the desired output is $(\text{out})_2=\ket\psi_2$. The $T_{2\leftarrow1}$ teleportation operator can be simplified and written effectively as\footnote{Note that $\theta=\tfrac{\pi}{2}$, i.e., performing a homodyne measurement of $Q$, does not teleport. It acts to disconnect the cluster state.}
\begin{equation}
\label{eq:Kraus_rot_homodyne}
    T_{2\leftarrow1}=X_2(\tfrac{m}{\cos\theta})R_2(\tfrac{\pi}{2})\mathcal{P}_2(\tan\theta)\text{SWAP}_{2\leftarrow1},
\end{equation}
where $\mathcal{P}(\kappa)=e^{i\kappa Q^2}$ is the shear operator, $X(s)=e^{-isP}$ and $Z(t)=e^{itQ}$ are the respective $Q$ and $P$-quadrature translation (a.k.a.\ displacement) operators, and the quantum Fourier transform $R_2(\tfrac{\pi}{2})$ is a teleportation byproduct---analog to the Hadamard gate for qubits. Although the $T_{2\leftarrow1}$ teleportation operator is applied by physically performing a destructive measurement on mode 1, the quantum information is preserved and teleported to mode 2, so one can treat $T_{2\leftarrow1}$ as an effective single-mode (one-in, one-out) operator. Another important point is that while the translation operator applied is dependent upon the randomly determined measurement result $m$, deterministic operation is always possible by the feedforward of $m$ to further qumodes, by shifting and rotating subsequent measurement bases. In fact, for Gaussian operations, this `adjustment' can even be made \textit{after} performing subsequent measurements by performing the correction on the measurement result~\cite{Menicucci2006}. This is because Gaussian CV gates are the analogues of Clifford qubit gates. Keeping this key fact in mind, we can treat \eq{canon_telep} as the operator
\begin{equation}
    T(\kappa)=R(\tfrac{\pi}{2})\mathcal{P}(\kappa),
\end{equation}
where $\kappa=\tan\theta$ is chosen by the user. 

Concatenating several circuits of this form is equivalent to teleporting down a cluster state chain of the form of \eq C, and one can show that any arbitrary single-mode Gaussian unitary can be implemented with four teleportation steps~\cite{Ukai2010}:
\begin{equation}
    {U_G}=T(\kappa_4)T(\kappa_3)T(\kappa_2)T(\kappa_1).
\end{equation}
Note that performing two consecutive measurements with $\kappa_2=\kappa_3=0$ allows for the transference of the $C_Z$ operation and effectively applies the two-mode gate ${CZ}_{14}$. A more detail and graphical understanding of this effect can be found in Ref.~\cite{Gu2009}. With the ability to apply ${CZ}_{ij}$ and single-mode Gaussian unitaries, one can thus apply arbitrary two-mode Gaussian unitaries through series of not more than 10 homodyne measurements. With homodyne measurement alone, one can thus use a cluster state to enact any multi-mode Gaussian unitary, which is the equivalent of the Clifford gate set for qubit-based computation. 

As is well known, the universal gate set for quantum computing requires at least one non-Clifford, i.e., one non-Gaussian, gate, by virtue of the CV translation~\cite{Bartlett2002,Mari2012} of the Gottesman-Knill theorem~\cite{Gottesman1999a}. It is also well known that the universal gate set can be derived from solely Gaussian measurement-based gates~\cite{Baragiola2019} when these are  applied to qumodes encoded in qubits by way of the Gottesman-Kitaev-Preskill (GKP) code~\cite{Gottesman2001}. An efficient protocol for the near-deterministic generation of GKP qubit states was discovered recently~\cite{Eaton2022_PhANTM} that fully leverages the CV cluster state structure presented above with the sole addition of PNR measurements to homodyne ones. 

Finally, a critical question is that of finite squeezing: quadrature eigenstates are unphysical and Gaussian squeezed states must be used in practice. The question is that the finiteness of the squeezing parameter $r$ introduces Gaussian noise which will ruin the CV protocol~\cite{Ohliger2010}. However, the GKP qubit encoding remedies that problem by allowing quantum error correction for small shifts $p\mapsto p+\varepsilon$ in the quadrature basis $\ket p$ and Menicucci proved the existence of a fault tolerance threshold for CVQC~\cite{Menicucci2014ft} requiring no more than $10\log[\exp(-2r)]=20.5$ dB of squeezing. This threshold was brought down to less than 10 dB~\cite{Fukui2018} by use of topological surface error code~\cite{Kitaev2003} foliated into a cluster state~\cite{Raussendorf2006,Raussendorf2007}.

\end{document}